\documentclass[prd,reprint,aps,amsmath,amssymb,showpacs,nofootinbib,
preprintnumbers,superscriptaddress]{revtex4-1}
\usepackage[utf8x]{inputenc}
\usepackage{graphicx}
\usepackage{enumitem}
\usepackage{slashed}
\usepackage{url}
\usepackage[pdftex]{hyperref}
\usepackage[T1]{fontenc}
\usepackage{bm}
\usepackage{color}
\usepackage{booktabs}
\usepackage{multirow}

\newcommand{\bra}[1]{\langle #1|}
\newcommand{\ket}[1]{|#1 \rangle}

\begin{document}

\preprint{PITT-PACC-1709}

\title{Baryogenesis from Oscillations of Charmed or Beautiful Baryons}
\author{Kyle Aitken}
\email{kaitken17@gmail.com}
\affiliation{Department of Physics, University of Washington, Seattle, WA
98195, USA}

\author{David McKeen}
\email{dmckeen@pitt.edu}
\affiliation{Pittsburgh Particle Physics, Astrophysics, and Cosmology Center,
Department of Physics and Astronomy, University of Pittsburgh, PA 15260, USA}

\author{Ann E. Nelson}
\email{aenelson@uw.edu}
\affiliation{Department of Physics, University of Washington, Seattle, WA
98195, USA}

\author{Thomas Neder}
\email{neder@ific.uv.es}
\affiliation{AHEP  Group,  Instituto  de  F\'{\i}sica  Corpuscular ---
C.S.I.C./Universitat  de  Val\`{e}ncia, Parc  Cient\'{\i}fic  de  Paterna, C/
Catedr\'{a}tico  Jos\'{e}  Beltr\'{a}n  2  E-46980  Paterna  (Valencia), Spain}
\date{\today}

\begin{abstract}
We propose a model for CP violating oscillations of neutral, heavy-flavored
baryons into antibaryons at rates which are within a few orders of magnitude of
their lifetimes. The flavor structure of the baryon violation suppresses
neutron oscillations and baryon number violating nuclear decays to
experimentally allowed rates. We also propose a scenario for producing such
baryons in the early Universe via the out-of-equilibrium decays of a neutral
particle, after hadronization but before nucleosynthesis. We find parameters
where CP violating baryon oscillations at a temperature of a few MeV could
result in the observed asymmetry between baryons and antibaryons. Furthermore,
part of the relevant parameter space for baryogenesis is potentially testable
at Belle II via decays of heavy flavor baryons into an exotic neutral fermion.
The model introduces four new particles: three light Majorana fermions and a
colored scalar. The lightest of these fermions is typically long lived on
collider timescales and may be produced in decays of bottom and possibly
charmed hadrons.
\end{abstract}


\maketitle

\section{Introduction}
\label{sec:intro}
The puzzle of \textit{baryogenesis}, how the Universe came to be composed
primarily of matter rather than equal amounts of matter and antimatter, has led
to numerous theories about physics beyond the standard model (SM), beginning
with the pioneering work of Sakharov~\cite{Sakharov:1967dj}. Three ingredients
are present in one form or another in any baryogenesis theory: baryon number
violation, C and CP violation, and departure from thermal equilibrium. Because
baryon number violation is required, initially baryogenesis was thought to
involve new baryon number violating processes which are only important at very
high energies, although it was later realized that anomalous electroweak
processes could do the job at temperatures as low as the weak phase
transition~\cite{Klinkhamer:1984di, *Kuzmin:1985mm, *Arnold:1987mh,
*Arnold:1987zg}.

Most baryogenesis models require the Universe to reheat after inflation to a
high temperature, typically well above the weak scale. However, many theories
of physics beyond the SM are inconsistent with a high inflation scale or are
inconsistent with a high postinflation reheat scale. Axion dark matter, if the
axion is present during inflation, requires a low inflation scale in order to
avoid excessive isocurvature perturbations~\cite{Turner:1983sj, *Seckel:1985tj,
*Turner:1990uz, *Fox:2004kb, *Beltran:2006sq, *Marsh:2014qoa}. Supersymmetry
requires a low reheat scale in order to avoid overproduction of the
gravitino~\cite{Moroi:1993mb, *Kawasaki:1994af, *Bolz:2000fu, *Kawasaki:2004qu,
*Kohri:2005wn, *Kawasaki:2006hm, *Kawasaki:2008qe}. The relaxion solution to
the hierarchy problem requires a low inflation scale so that the Hubble
temperature during inflation does not suppress
instantons~\cite{Graham:2015cka}. In addition, avoiding the need for a high
reheat temperature or production of heavy particles during reheating means a
low baryogenesis scale is consistent with a wider variety of inflationary
models~\cite{Kofman:1994rk, *Shtanov:1994ce}.

Lower reheat temperatures are possible provided the inflaton decays produce
heavy particles which decay out of thermal equilibrium in a baryon and CP
violating manner~\cite{Dimopoulos:1987rk}. In Ref.~\cite{Dimopoulos:1987rk}
baryogenesis occurs due to the baryon number violating decays of TeV mass
squarks in an R-parity violating supersymmetric model, in which the reheat
temperature could be as low as an MeV, provided that the heavy squarks can be
produced out of equilibrium at the end of inflation. Such squark mediated
baryon number violation is consistent with the observed lifetime of the proton,
due to the conservation of lepton number, and, depending on the flavor
structure of the baryon number violating operators, can be consistent with the
stability of heavy nuclei as well. In Ref.~\cite{Kuzmin:1996sy} it was pointed
out that such heavy flavor baryon number violation could lead to oscillations
of the $\Xi_{b}^{0}$ baryon at a rate comparable to its lifetime, while being
consistent with the lifetime of heavy nuclei.

Here we present a baryogenesis model which is consistent with a reheat
temperature as low as a few MeV, and which requires no postinflationary
production of any particle heavier than about 6---10 GeV. The required baryon
number violation is conceivably observable via the oscillations of heavy-flavor
neutral baryons, and the required CP violation is potentially of ${\cal O}(1)$
in such oscillations. The processes that produce the baryon asymmetry in the
early Universe involve particles and phenomena which can be directly studied in
the laboratory -- a unique feature of our theory. Our proposal is that certain
neutral heavy flavor baryons undergo CP and baryon number violating
oscillations and decays, and are produced in the early Universe via the out of
equilibrium decays of a weakly coupled neutral particle whose lifetime is of
order $0.1~\rm s$, a time when the temperature is of order a few MeV. The
basic scenario was outlined in Ref.~\cite{McKeen:2015cuz}, and the model we
study has the same field content and couplings as Ref.~\cite{Ghalsasi:2015mxa}.
The basic formalism for analyzing CP violation in fermion antifermion
oscillations was worked out in Ref.~\cite{Ipek:2014moa}.

The outline of the paper is as follows. In section \ref{sec:model} the model is
introduced, and the effective operator responsible for baryon oscillations is
constructed. In section \ref{sec:Heavy_Flavor_Baryon_Osc}, general $\Delta
B=2$, six-quark effective operators are analyzed for their contribution to
dinucleon decay, that is, the decay of two nucleons into mesons. Currently,
dinucleon decay places similar or stronger constraints on all such operators
than does neutron oscillations. For operators that cannot contribute to
dinucleon decay at tree level, electroweak corrections to the six-quark
operators are examined. In section \ref{heavyBosc}, the general formalism for
CP-violating oscillations of fermions is reviewed, and the oscillation
parameters are calculated for the model introduced in section \ref{sec:model}.
In sections \ref{sec:colliders} and \ref{sec:hadrons}, direct constraints on
the masses and couplings of the new $\phi$ and $\chi$ particles from collider
searches, and indirect constraints from rare decays of mesons and baryons are
derived, respectively. Section \ref{sec:cosmo} contains our analysis of how in
this model the baryon asymmetry of the Universe (BAU) is produced. Finally, in
section \ref{sec:summary} we conclude and point at possible directions for
future work.

\section{Model}
\label{sec:model}
We wish to find a theory which allows for sufficiently large baryon number and
CP violation to explain baryogenesis at relatively low energy. In order to
ensure sufficient stability of the proton, we assume lepton number is not
violated, other than perhaps via the tiny $\Delta L=2$ terms that could account
for Majorana neutrino masses. The lowest dimension terms which violate baryon
number and not lepton number are dimension 9, six-quark $\Delta B=2$ operators.
Such operators can lead to neutral baryon oscillations and conceivably CP
violation~\cite{McKeen:2015cuz}, and can arise as an effective field theory
description of physics at some higher energy scale. A minimal renormalizable
model for generating such terms involves a new charge $-1/3$ color triplet
scalar and two Majorana fermions, as described in Ref.~\cite{Ghalsasi:2015mxa}.
A third Majorana fermion, which decays out of thermal equilibrium, allows for
the fulfillment of the out-of-equilibrium Sakharov condition.

We note that this model for baryon number violation can easily be embedded in
an $R$-parity violating supersymmetric (RPV SUSY) theory. In such theories, the
neutralinos would play the role of the Majorana fermions and a down-type SU(2)
singlet squark can be the colored scalar. For simplicity, we do not explore
this embedding in a SUSY framework in this paper and we stick to the minimal
version of the model.

Our model thus adds four new particles: three Majorana fermions,
$\chi_{1,2,3}$, and a single color triplet scalar, $\phi$. The interactions
involving the new particles and weak SU(2) singlet SM quarks are given by
\begin{equation}
{\cal L}_{\rm int}\supset -g_{ud}\phi^\ast \bar{u}_R d_R^c-y_{id}\phi\bar \chi_i d_R^c+{\rm h.c.},
\label{eq:L_int}
\end{equation}
along with terms involving other generations, $d\to s,b$ and $u\to c,t$. By
convention we take all two component fields to transform in the left-handed
representation under Lorentz transformations. $d_R^c$ stands for the charge
conjugate of the right-handed down quark field, which is in the left-handed
Lorentz representation. 

The required new particles and their interactions are motivated as follows. A
natural way to construct the $\Delta B=2$ six-quark operator we require for
baryon oscillations is from two $\Delta B=1$ four-fermion interactions
connected by an exotic neutral Majorana fermion. Thus we introduce an exotic,
electrically-neutral, colorless, Majorana fermion, $\chi_{1}$, which couples to
other fermions via a four-fermion interaction of the form $u_Rd_Rd_R'\chi_1$
(using $u$ and $d$ here to represent any up- or down-type quark).

Since such a four-fermion interaction is itself nonrenormalizable, we also
introduce a complex, color triplet, scalar particle (diquark) $\phi$ to mediate
the $\Delta B=1$ interactions. Note that if $\chi_1$ is heavier than the
difference in mass between the proton and electron, $m_p-m_e=937.76~\rm MeV$,
this interaction does not give rise to proton decay.\footnote{The stability of
$^9$Be leads to a marginally stronger lower bound of $m_{\chi_1}>937.9~\rm
MeV$~\cite{McKeen:2015cuz}.} In the presence of only $\chi_1$, there is no
physical CP violation, as there is enough reparameterization freedom to remove
the phases in the couplings. We introduce a second fermion, $\chi_2$ (with
$m_{\chi_2}>m_{\chi_1}$), in order to give rise to CP violation. Finally, for
baryogenesis, the oscillating baryons must be produced out of thermal
equilibrium. As described in Sec. \ref{sec:cosmo}, this is most simply
accomplished by introducing a third Majorana fermion, $\chi_{3}$, which decays
out of equilibrium to produce the baryons whose oscillations result in
baryogenesis.

Note that we only consider operators constructed out of right-handed quarks,
for two reasons. Our phenomenological reason is that, as we will show in
Sec.~\ref{sec:Heavy_Flavor_Baryon_Osc}, right handed quark operators are less
constrained by dinucleon decay due to the requirement of light quark mass
insertions in flavor-changing loops. Our top down theoretical reason is that,
as mentioned, the interactions in Eq.~(\ref{eq:L_int}) occur in RPV SUSY
models, suggesting a possible embedding of our model into a more complete
theory.

\begin{figure}
\includegraphics[width=\linewidth]{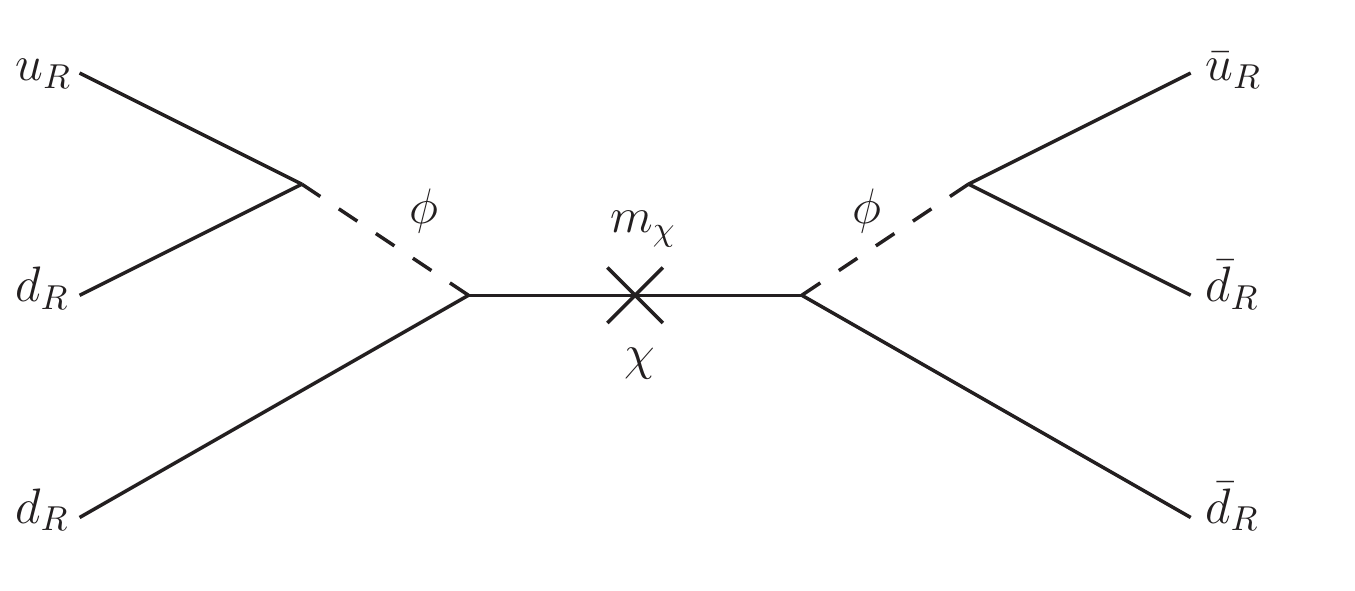}
\caption{The basic six-quark $\Delta B=2$ operator generated by $\phi$ and $\chi$ exchange. $u$ and $d$ here represent any of the up- or down-type quark flavors. The quarks involved are all weak SU(2) singlets, as emphasized by the $R$ subscripts.}
\label{fig:six quark_chi_phi}
\end{figure}
Figure~\ref{fig:six quark_chi_phi} shows the $\Delta B=2$ six-quark operators
that are generated by the interactions in Eq.~(\ref{eq:L_int}). Such operators,
which can mediate the transition of a baryon $\cal B$ to an antibaryon
$\bar{\cal B}^\prime$, can be written as 
\begin{equation}
\begin{aligned}
&{\cal O}_{\cal B{\cal B}^\prime}=\epsilon^{abc}\epsilon^{def}
\\
&~\times\left[\left(q_R\right)^i_a\left(q_R\right)_{i,b}\left(q_R\right)^j_c\left(q_R^\prime\right)_{j,d}\left(q_R^\prime\right)^k_e\left(q_R^\prime\right)_{k,f}+\dots\right],
\label{eq:op_six_quark}
\end{aligned}
\end{equation}
where $a,\dots,f$ are color indices, $i,j,k$ spinor indices, and $q=u,d,s,c,b$
any of the quark flavors (because of its short lifetime, the top quark does not
hadronize and is not important in the low-energy effective theory) which are
all right-chiral. The ellipsis represents other possible permutations of color
or spinor indices. Here, ${\cal B}$ denotes an arbitrary standard model baryon
with the quark content $qqq$ while ${\cal B}^\prime$ contains $q^\prime
q^\prime q^\prime$. [We will use both the baryon name $\cal B$ or the quark
content $(qqq)$ to label the operators in question throughout this paper.] The
precise index structure of ${\cal O}_{\cal B{\cal B}^\prime}$ is not important
for the purposes of this paper. Therefore, in what follows, we will suppress
the indices on ${\cal O}_{\cal B{\cal B}^\prime}$ and generically denote the
operators we are interested in that appear in the effective Lagrangian via the
shorthand
\begin{equation}
{\cal L}_{\rm eff}\supset C_{\cal B{\cal B}^\prime}\left(qqq\right)\left(q^\prime q^\prime q^\prime\right)\equiv C_{\cal B{\cal B}^\prime}{\cal O}_{\cal B{\cal B}^\prime},
\label{eq:LeffBBprime}
\end{equation}
keeping in mind that the leading operators that are generated involve only
right-chiral quarks.

Matching the interactions generated by Eq.~(\ref{eq:L_int}) to the effective
theory at tree-level gives the coefficient of the operator that generates
oscillations between a neutral baryon and its antiparticle, $\cal
B\leftrightarrow\bar{\cal B}$,
\begin{equation}
C_{\cal BB}\sim\sum_{i}\frac{{m_{\chi}}_i}{m_{\cal B}^2-{m_{\chi}}_i^2}\left(\frac{g_{ud}^\ast y_{id^\prime}+g_{ud^\prime}^\ast y_{id}}{m_\phi^2}\right)^2,
\label{eq:CBB}
\end{equation}
with $u$, $d$, and $d^\prime$ labeling the quarks comprising $\cal B$. For
example, the operator $\left(ddc\right)^2$ would allow the processes
$\bar{\Sigma}_{c}\leftrightarrow\Sigma_{c}$. Given this operator, we will find
it useful to relate the coefficient to the (dispersive) transition amplitude,
defined by $\delta_{\cal BB}\equiv\bra{\bar{\cal B}}C_{\cal BB}{\cal O}_{\cal
BB}\ket{{\cal B}}$, with
\begin{equation}
\delta_{\cal BB}=\kappa^2C_{\cal BB},
\label{eq:delta}
\end{equation}
where $\kappa\sim10^{-2}~\rm GeV^3$~\cite{Buchoff:2015qwa}. In analogy with
meson oscillations, when the two-state system in question is unambiguous,
$\delta_{\cal BB}$ can also be referred to as $M_{12}$.

Operators which involve different baryons of the form ${\cal O}_{\cal
BB^\prime}$ would allow for a common decay product between $\cal B$ and
$\bar{\cal B}$ and could also give rise to oscillations. For example,
$\left(uss\right)\left(uds\right)$ would allow for $\Xi^{0}$ and
$\bar{\Xi}^{0}$ to have a common decay product, through $\Xi^{0}\to\Lambda^{0}$
and $\bar{\Xi}^{0}\to\Lambda^{0}$ (ignoring any neutral meson products).
However, we will find such processes are suppressed relative to their direct
oscillation cousins, so we will ignore them in our analysis. There are also
baryon-number--preserving operators that contribute to the masses and mixings
of SM baryons. These are greatly suppressed relative to those that occur in the
SM and we do not consider them further either. Therefore, in what follows, we
focus on operators ${\cal O}_{\cal BB}$ with coefficients of the form of
Eq.~(\ref{eq:CBB}).

\section{Dinucleon Decay Constraints
\label{sec:Heavy_Flavor_Baryon_Osc}}
As described in the preceding section, we would like our six-quark operators to
allow for the oscillation of heavy baryons in order to produce the Universe's
observed baryon asymmetry. In Sec.~\ref{sec:cosmo}, we will show that the ideal
width for such an oscillation, which is dependent upon the value of $C_{\cal
B\cal B}$ in Eq.~(\ref{eq:CBB}), is a few orders of magnitude smaller than
$\cal B$'s decay width. However, models with $B$ violation by two units are
certainly not a new idea, and so significant experimental effort has been put
fourth into constraining $\Delta B=2$ processes.  The most immediate constraint
on our six-quark operators is the lack of observed dinucleon decay, which we
quantify in this section.  The analysis we perform here applies to six-quark
operators in general and is independent of the origin of the new physics
introduced in Sec.~\ref{sec:model}.

Dinucleon constraints come from underground detectors whose primary purpose is
the detection of proton decay and neutrino oscillations. For example, in a
nucleus, a $n\to\bar n$ transition will be shortly followed by the annihilation
of the $\bar n$ with one of the other nucleons, leading to the decay of the
nucleus of mass number $A$ to a nucleus with $A^\prime=A-2$ plus mesons. The
lack of observation of such decays can therefore bound the transition amplitude
$\delta _{nn}$~\cite{Friedman:2008es} which is related to the coefficient of
the $(udd)^2$ operator, $C_{nn}$ [cf. Eqs.~(\ref{eq:CBB}) and
(\ref{eq:delta})]. Currently, the lower bound on the $^{16}O$ lifetime (in
decays to pions) of $1.9\times10^{32}~\rm years$ from the Super-Kamiokande
collaboration~\cite{Abe:2011ky} places the strongest limit, $\delta
_{nn}<1.9\times10^{-33}~\rm GeV$.

Operators that also violate strangeness do not directly induce $n\to\bar n$
transitions in a nucleus. However, they can also lead to dinucleon decays,
$A\to(A-2)+\rm mesons$, through the reaction $NN\to{\rm kaons}+X$ where $N$ is
a nucleon. 
\begin{figure}
\raisebox{3mm}[0pt][0pt]{\includegraphics[width=0.48\linewidth]{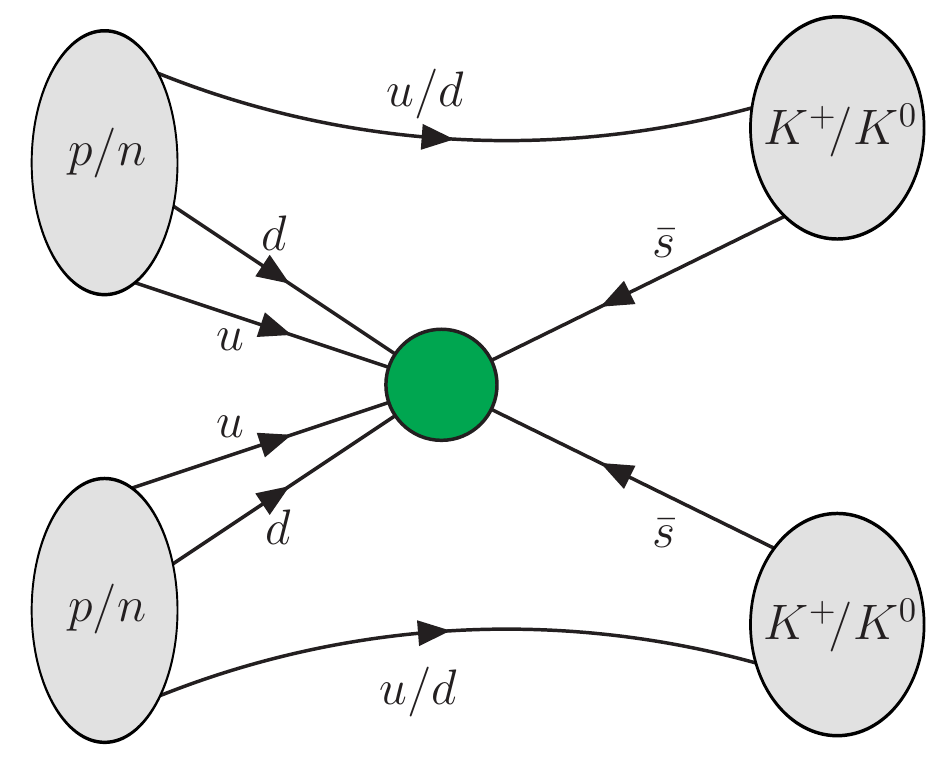}}\quad\includegraphics[width=0.48\linewidth]{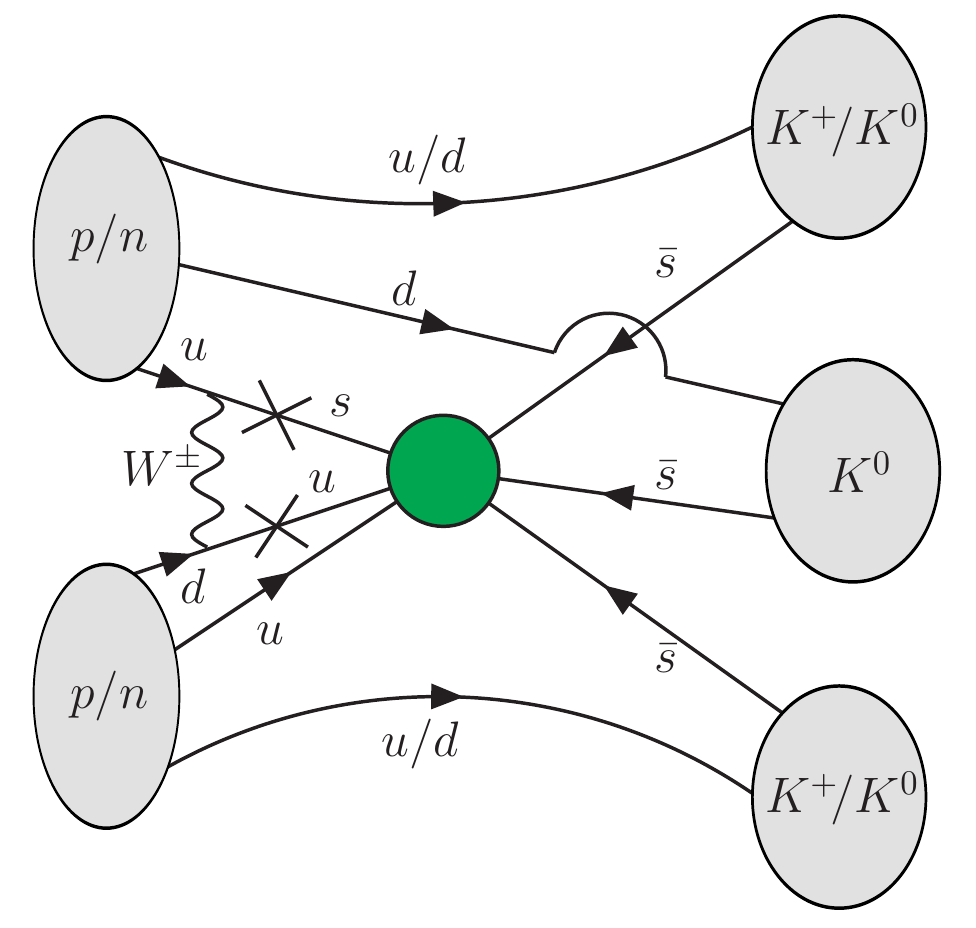}
\caption{Left: Dinucleon decay via the $\Delta B=\Delta S=2$ $(uds)^2$ operator
that mediates $\Lambda^0\leftrightarrow\bar\Lambda^0$ oscillations. Right:
Dinucleon decay mediated by the $\Delta B=2$, $\Delta S=4$ $(uss)^2$ operator
that becomes $\Delta S=3$ $(uds)(uss)$ operator in the presence of
flavor-changing weak interactions. Because the short-distance $\Delta B=2$
operators we consider involve weak isosinglets, this operator requires light
quark chirality flips, indicated by crosses. See text for discussion of the
matching of the short distance theory onto the (chiral symmetry violating) long
distance theory.}
\label{fig:DND_deltaS}
\end{figure}
For example, the diagram on the left of Fig.~\ref{fig:DND_deltaS} shows how the
operator $(uds)^2$ can lead to dinucleon decay to a pair of kaons. The
Super-Kamiokande collaboration~\cite{Litos:2014fxa} has searched for such
decays and has placed an upper bound on the $pp\to K^{+}K^{+}$ decay rate by
limiting the lifetime for $^{16}O\to{^{14}}C\, K^{+}K^{+}$ to more than
$1.7\times10^{32}~\rm years$.

To make use of this limit, we start with the effective operator ${\cal O}_{\cal
BB}$. The dinucleon decay rate through direct nucleon annihilation can then be
roughly approximated by considering the decay rate to a meson
pair~\cite{Goity:1994dq},
\begin{equation}
\Gamma_{NN\to\rm X}\sim\frac{9}{32\pi}\frac{\left|C_{\cal BB}\right|^2}{m_N^2}\left|\bra{2~{\rm mesons}}{\cal O}_{\cal BB}\ket{NN}\right|^2\rho_N
\label{eq:DND_width}
\end{equation}
where $m_N$ is the nucleon mass, $\rho_N\simeq0.25~\rm fm^{-3}$ is the nucleon
density, and we have ignored the masses of the final state particles. In the
case of operators that can contribute at tree level, the matrix element can be
estimated as roughly $\bra{2~{\rm mesons}}{\cal O}_{\cal
BB}\ket{NN}\sim\Lambda_{\rm QCD}^{5}\simeq\left(200~\rm MeV\right)^{5}$. Using
this and Eq.~(\ref{eq:delta}), the limit on the rate for $NN\to KK$ from Super
Kamiokande translates to a limit on the transition amplitude of
\begin{equation}
\delta_{(uds)^2}\lesssim 10^{-30}~\rm GeV.
\end{equation}
In what follows, we also take operators that change strangeness by one or three
units to have roughly the same bound as this.

Kinematic constraints protect certain operators from contributing to dinucleon
decay at leading order. Operators such as $(uss)^2$ that change strangeness by
four units (i.e. $\Delta S=4$) are kinematically forbidden from contributing to
dinucleon decay at tree level since $2m_N<4m_K$. Similarly, those that involve
charm\footnote{Depending on the nucleon binding energy, $nn\to D \gamma$
through a $\Delta C=1$ operator is kinematically allowed for some nuclei, but
due to the dependence of the amplitude on the photon momentum and coupling and
phase space suppression the rate is proportional to $(\alpha/4\pi)
\left(k_\gamma/m_N\right)^3\sim 10^{-9}$, where $k_\gamma$ is the photon
energy, suppressing the rate below other decays with less constrained phase
space.} or bottom quarks also do not lead to dinucleon decay at leading order.
However, when combined with flavor-violating weak interactions, these operators
involving heavy quarks can lead to dinucleon decay. An illustration of this is
shown on the right of Fig.~\ref{fig:DND_deltaS}. 

To properly estimate the rate for dinucleon decay from $\Delta B=2$ operators
(involving heavy flavors), we must match the UV theory involving quarks to a
low energy effective theory involving baryons valid at momentum transfers below
$4\pi f_\pi\sim 1~\rm GeV$ where $f_\pi=93~\rm MeV$. This consists of writing
down an operator in the UV theory and treating the coefficient of this operator
as a spurion that transforms in a particular way under the global chiral quark
flavor symmetry ${\rm SU}(3)_L\times {\rm SU}(3)_R$ so as to make the operator
invariant. This operator is then matched onto an operator in the effective
theory that transforms in the same way under the chiral symmetry with the same
spurion coefficient. In the UV, the light quarks $q_{L,R}$ transform as
triplets under ${\rm SU}(3)_{L,R}$. In the low energy theory, the meson octet,
$\Pi$, is described by a field $\Sigma=\exp\left(2i\Pi/f_\pi\right)$ which
transforms under the chiral symmetry as $\Sigma\to L\Sigma R^\dag$ where $L$,
$R$ are ${\rm SU}(3)_{L,R}$ transformations, respectively. Incorporating the
baryon octet (see, e.g., Ref.~\cite{Georgi:1985kw}) can be done by defining a
field $\xi=\exp\left(i\Pi/f_\pi\right)$ which transforms as $\xi\to L\xi
U^\dag$, $\xi\to U\xi R^\dag$ under ${\rm SU}(3)_{L,R}$. $U$ is an ${\rm
SU}(3)$ matrix that depends nonlinearly on the meson fields. The baryon octet
$B$ is defined to transform as $B\to UBU^\dag$. Operators in the effective
theory are then constructed out of $\Sigma$, $B$, and $\xi$ along with spurions
from the UV theory to be invariant under the flavor symmetry. Since the chiral
symmetry is dynamically broken by the strong coupling of QCD around $4\pi
f_\pi$, one can use naive dimensional analysis to properly account for factors
of $4\pi$ (that come from the strong coupling) and the cutoff, $4\pi f_\pi$,
that appear in this matching procedure, as described in, e.g.,
Ref.~\cite{Cohen:1997rt}.

We will first illustrate this matching procedure in our theory with
interactions given by Eq.~(\ref{eq:L_int}), assuming for now that only the
light quarks $u$, $d$, and $s$ are involved. We will deal with heavy quarks $c$
and $b$ below. Since distinction between chiralities is necessary, we will
temporarily denote them explicitly. After integrating out the scalar, $\phi$,
and the Majorana fermions, $\chi_i$, we are left with a $\Delta B=2$ operator
involving only (light) right-chiral quarks, $C_{\cal B\cal B}(q_R q_R q_R)^2$.
This operator must be matched onto an operator valid at long distances
involving baryons at the scale of chiral symmetry breaking. The coefficient
$C_{\cal B\cal B}$ can be treated as a spurion that transforms under ${\rm
SU}(3)_R$ in a representation that appears in the tensor decomposition of 6
triplets. For definiteness, take it to transform as an ${\rm SU}(3)_R$ octet.
Then the object $\tilde C_{\cal B\cal B}\equiv\xi C_{\cal B\cal B} \xi^\dag$
transforms as $\tilde C_{\cal B\cal B}\to U\tilde C_{\cal B\cal B} U^\dag$ and
the operator matching is
\begin{equation}
C_{\cal B\cal B}(q_R q_R q_R)^2\to \left(4\pi f_\pi^3\right)^2{\rm tr}B \tilde C_{\cal B\cal B} B+\dots,
\end{equation}
where the ellipsis represents other possible orderings of the baryon octets and
the spurion. Note that this gives an understanding of the size of $\kappa\sim
4\pi f_\pi^3\simeq 10^{-2}~\rm GeV^3$ in Eq.~(\ref{eq:delta}). Adjusting this
analysis if $C_{\cal B\cal B}$ transforms under a different representation of
${\rm SU}(3)_R$ is straightforward; one inserts the required numbers of $\xi$
and $\xi^\dag$ into the definition of $\tilde C_{\cal B\cal B}$ so that it
transforms in such a way as to leave ${\rm tr}B \tilde C_{\cal B\cal B} B$
invariant. For example, if $C_{\cal B\cal B}$ is a singlet then one simply
takes $\tilde C_{\cal B\cal B}\equiv C_{\cal B\cal B}$. 

Four-quark weak operators involving light quarks can be matched onto the low
energy effective theory in much the same way. The coefficient of the operator
$\bar u_L\gamma^\mu q_L^j \bar {q_L}_i\gamma_\mu u_L$ can be viewed as a
spurion that transforms as an octet under ${\rm SU}(3)_L$ and the strangeness
changing ($\Delta S=1$) coefficient takes a value $\propto G_FV_{us}V_{ud}^\ast
h$ with $h^i_{\, j}=\delta^i_2\delta^3_j$. Then $\xi^\dag h\xi\to U\xi^\dag
h\xi U^\dag$ and the matching is
\begin{equation}
\begin{aligned}
&\frac{G_F}{\sqrt{2}}V_{us}V_{ud}^\ast \bar u\gamma^\mu \left(1-\gamma^5\right)s \bar d\gamma_\mu \left(1-\gamma^5\right)u
\\
&\quad\to \frac{G_F}{\sqrt{2}}V_{us}V_{ud}^\ast \left(4\pi f_\pi^3\right) {\rm tr} \bar B \xi^\dag h\xi B+\dots,
\end{aligned}
\end{equation}
where again the ellipsis represents other possible orderings of $B$, $\bar B$,
and $\xi^\dag h\xi$. 

Now we can combine a $\Delta B=2$ operator that also changes strangeness by $n$
units with the weak $\Delta S=1$ operator to form a $\Delta B=2$, $\Delta
S=n-1$ operator that is given by
\begin{equation}
\frac{G_F}{\sqrt{2}}V_{us}V_{ud}^\ast f_\pi^2 \left(4\pi f_\pi^3\right)^2{\rm tr} \bar B \tilde C_{\cal B\cal B}\xi^\dag h\xi B+\dots.
\end{equation}
In other words, if the leading $\Delta B=2$ operator has $\Delta S=n$, the
$\Delta B=2$, $\Delta S=n-1$ operator that is generated due to weak
interactions is suppressed relative to it by the factor
\begin{equation}
\frac{G_F}{\sqrt{2}}V_{us}V_{ud}^\ast f_\pi^2\sim 10^{-8}.
\end{equation}
Thus, for example, the bound on a leading $\Delta S=4$ operator $(u_Rs_Rs_R)^2$
from the lack of dinucleon decay is around eight orders of magnitude weaker
than that on the $\Delta S=3$ operator $(uds)(uss)$ [which we take to be
comparable to that on the $\Delta S=2$ operator $(uds)^2$],
\begin{equation}
\delta_{(uss)^2}\lesssim 10^{-22}~\rm GeV.
\end{equation}

\begin{figure}
\includegraphics[width=\linewidth]{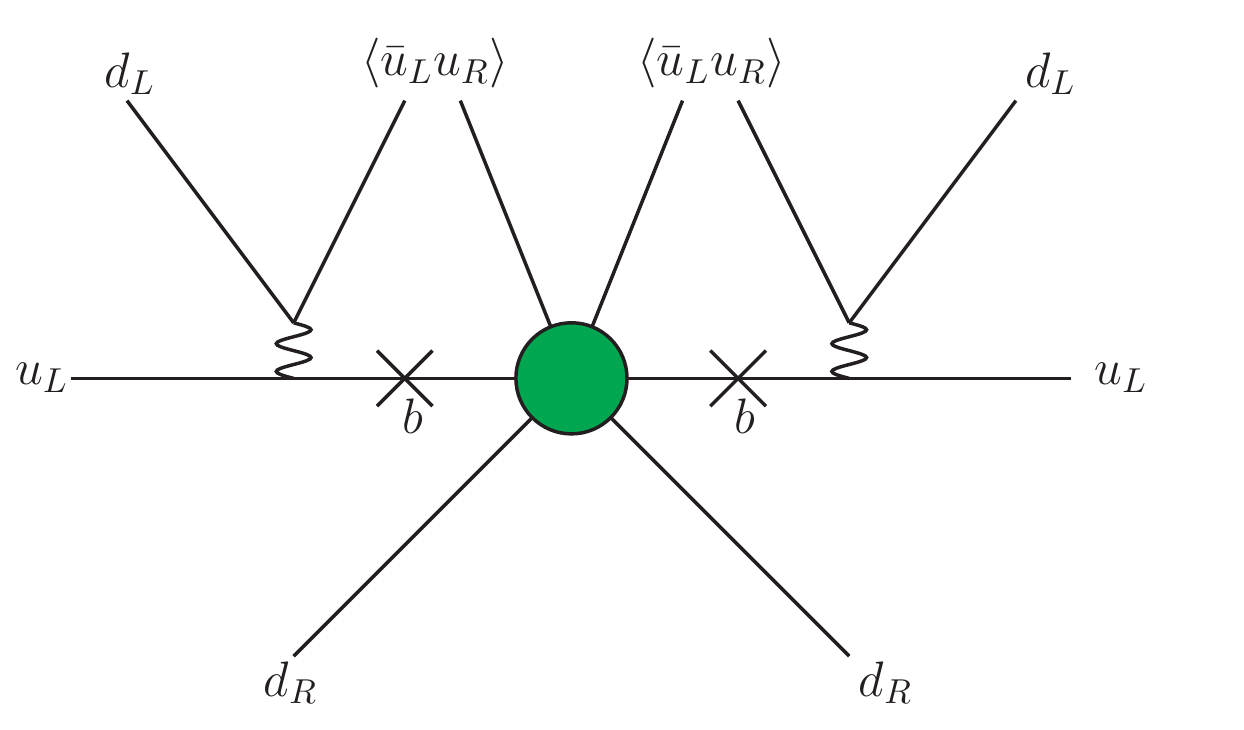}
\caption{The ten-quark $\Delta B=2$ operator that results from the leading
$(u_Rd_Rb_R)^2$ operator after integrating out the $b$ quarks. The crosses
represent chirality-flipping $b$ quark mass insertions. We use $\langle\bar u_L
u_R\rangle$ to indicate the pairs of light quark fields that can be replaced by
the chiral condensate when matching onto the long distance theory relevant for
dinucleon decay.}
\label{fig:DND_deltaB}
\end{figure}

Now, we consider the case where the leading $\Delta B=2$ operators contain
heavy quarks. Consider, for example, if after integrating out the heavy scalar
$\phi$ and Majorana fermions $\chi_i$, that the leading operator we generate is
$C_{(udb)^2}(u_Rd_Rb_R)^2$. Before matching onto the theory valid after chiral
symmetry breaking we must first integrate out the $b$ quarks. In the presence
of weak interactions, as shown in Fig.~\ref{fig:DND_deltaB}, doing so will lead
to a ten-quark operator,
\begin{equation}
\begin{aligned}
C_{(udb)^2}(u_Rd_Rb_R)^2&\to \left(\frac{G_F}{\sqrt{2}}V_{ub}V_{ud}^\ast\frac{1}{m_b}\right)^2
\\
&\quad\quad\times C_{(udb)^2}(u_Rd_Rd_L\bar u_L u_L)^2.
\end{aligned}
\end{equation}
After chiral symmetry breaking, $\bar u_L u_R$ can be replaced by the quark
condensate which is roughly $4\pi f_\pi^3$. This means that the induced $\Delta
B=2$ operator $(u_Ld_Rd_L)^2$ is suppressed relative to $(u_Rd_Rb_R)^2$ by the
factor
\begin{equation}
\left(\frac{G_F}{\sqrt{2}}V_{ub}V_{ud}^\ast\frac{4\pi f_\pi^3}{m_b}\right)^2 \sim 10^{-20}.
\end{equation}
Additionally in the case of a leading operator containing a $b$ and $c$ quark,
e.g. $(d_Rc_Rb_R)^2$, there are perturbative loops that generate dimension-nine
operators involving light quarks above the chiral symmetry breaking scale. In
the case of the operator $(d_Rc_Rb_R)^2$, two such loops can be used to
generate the operator $(u_Ld_Ld_R)^2$ with a coefficient suppressed relative to
the leading one by
\begin{equation}
\left(\frac{G_F}{\sqrt{2}}V_{ub}V_{cd}^\ast \frac{m_b m_c}{4\pi^2}\log\frac{m_W^2}{m_b^2}\right)^2 \sim 10^{-16}.
\end{equation}

In Table~\ref{tab:operators}, we list operators that can mediate ${\cal
B}\leftrightarrow\bar{\cal B}$ transitions along with the number of loops
required for each operator to mediate ($\Delta S=0,1,2,3$) dinucleon decay. We
show the resulting limits on the transition amplitudes $\delta_{{\cal B}{\cal
B}}=\left|M_{12}\right|=\kappa^2 C_{{\cal B}{\cal B}}$ of each operator from
the lack of observation of dinucleon decay, accounting for the appropriate
suppression factors. In general we find that only operators which require 2 or
more weak interactions to contribute to dinucleon decay can give baryon
oscillations at a rate which is large enough to be relevant for either
experimental searches or baryogenesis. The last column of the table gives the
limit on the size of the operator that can be produced in our specific model
when collider constraints on new particles are considered, which will be
discussed in Sec.~\ref{sec:colliders}.
\begin{table*}[t]
\begin{tabular}{cccccc}
\hline\hline
\multirow{2}*{Operator} & \multirow{2}*{${\cal B}$} & Weak Insertions & Measured & \multicolumn{2}{c}{Limits on $\delta_{\cal BB}=M_{12}~$($\rm GeV$)} \\
 & & Required & $\Gamma$ ($\rm GeV$)~\cite{Olive:2016xmw} & Dinucleon decay & Collider \\
\hline\hline
$(udd)^2$ & $n$ & None & $\left(7.477\pm0.009\right)\times10^{-28}$ & $10^{-33}$ & $10^{-17}$\\
$(uds)^2$ & $\Lambda$ & None & $\left(2.501\pm0.019\right)\times10^{-15}$ & $10^{-30}$ & $10^{-17}$\\
$(uds)^2$ & $\Sigma^0$ & None & $\left(8.9\pm0.8\right)\times10^{-6}$ & $10^{-30}$ & $10^{-17}$\\
$(uss)^2$ & $\Xi^0$ & One & $\left(2.27\pm0.07\right)\times10^{-15}$ & $10^{-22}$ & $10^{-17}$\\
$(ddc)^2$ & $\Sigma_c^0$ & Two & $\left(1.83^{+0.11}_{-0.19}\right)\times10^{-3}$ & $10^{-17}$ & $10^{-16}$\\
$(dsc)^2$ & $\Xi_c^0$ & Two & $\left(5.87^{+0.58}_{-0.61}\right)\times10^{-12}$ & $10^{-16}$ & $10^{-15}$\\
$(ssc)^2$ & $\Omega_c^0$ & Two & $\left(9.5\pm1.2\right)\times10^{-12}$ & $10^{-14}$ & $10^{-15}$\\
$(udb)^2$ & $\Lambda_b^0$ & Two & $\left(4.490\pm0.031\right)\times10^{-13}$ & $10^{-13}$ & $10^{-17}$\\
$(udb)^2$ & ${\Sigma_b^0}^\ast$ & Two & $\sim {10^{-3}}^\ast$ & $10^{-13}$ & $10^{-17}$\\
$(usb)^2$ & $\Xi_b^0$ & Two & $\left(4.496\pm0.095\right)\times10^{-13}$ & $10^{-10}$ & $10^{-17}$\\
$(dcb)^2$ & ${\Xi_{cb}^0}^\dag$ & Two & $\sim {10^{-12}}^\dag$ & $10^{-17}$ & $10^{-15}$\\
$(scb)^2$ & ${\Omega_{cb}^0}^\dag$ & Two & $\sim {10^{-12}}^\dag$ & $10^{-14}$ & $10^{-15}$\\
$(ubb)^2$ & ${\Xi_{bb}^0}^\ddag$ & Four & $\sim {10^{-13}}^\ddag$ & >1 & $10^{-17}$\\
$(cbb)^2$ & ${\Omega_{cbb}^0}^\dag$ & Four & $\sim {10^{-12}}^\dag$ & >1 & $10^{-15}$
\\\hline\hline
\end{tabular}
\caption{Operators that mediate ${\cal B}\leftrightarrow\bar{\cal B}$
oscillations and the number of weak interaction insertions required for each of
these to contribute to dinucleon decay. The resulting limit from dinucleon
decay on the transition amplitude, defined in Eq.~(\ref{eq:delta}), for each
operator is shown. An $\ast$ indicates a baryon that has not yet been observed
and which has a strong decay channel open. A $\dag$ ($\ddag$) indicates an
unobserved baryon which primarily decays through a weak interaction of a $c$
($b$) quark.}
\label{tab:operators}
\end{table*}

\section{CP Violation in Heavy Baryon Oscillations}
\label{heavyBosc}
The evolution of the $({\cal B},\bar{\cal B})$ system in vacuum, assuming CPT
conservation, can be described~\cite{McKeen:2015cuz} by a $2\times2$
Hamiltonian,
\begin{equation}
{\cal H}=M-\frac{i}{2}\Gamma=\left(\begin{array}{cc}
M_{\cal B}-\frac{i}{2}\Gamma_{\cal B} & M_{12}-\frac{i}{2}\Gamma_{12} \\
M_{12}^\ast-\frac{i}{2}\Gamma_{12}^\ast & M_{\cal B}-\frac{i}{2}\Gamma_{\cal B}
\end{array}\right).
\label{eq:ham}
\end{equation}
$M$ and $\Gamma$ are both Hermitian matrices that describe the dispersive and
absorptive parts of the ${\cal B},\bar{\cal B}\to{\cal B},\bar{\cal B}$
amplitude, respectively. This system is entirely analogous to the very well
known case of neutral mesons and antimesons. Because of the off-diagonal terms
in $\cal H$, the mass eigenstates $\ket{{\cal B}_{\rm L,H}}$ with masses
$m_{\rm L,H}$ are linear combinations of the flavor eigenstates $\ket{\cal B}$
and $\ket{\bar{\cal B}}$,
\begin{equation}
\ket{{\cal B}_{\rm L,H}}=p\ket{{\cal B}}\pm q\ket{\bar{\cal B}}.
\end{equation}
The mass difference is $\Delta m=m_{\rm H}-m_{\rm L}>0$ and the width
difference between the states is $\Delta\Gamma=\Gamma_{\rm H}-\Gamma_{\rm L}$
and can be of either sign. The flavor admixtures can be determined by
\begin{equation}
\left(\frac qp\right)^2=\frac{M_{12}^\ast-(i/2)\Gamma_{12}^\ast}{M_{12}-(i/2)\Gamma_{12}}.
\end{equation}
A state that begins at $t=0$ as a $\ket{\cal B}$ or $\ket{\bar{\cal B}}$ is at time $t$
\begin{equation}
\begin{aligned}
\ket{{\cal B}(t)}&=g_+(t)\ket{{\cal B}}-\frac qp g_-(t)\ket{\bar{\cal B}},
\\
\ket{\bar{\cal B}(t)}&=g_+(t)\ket{\bar{\cal B}}-\frac pq g_-(t)\ket{{\cal B}}
\end{aligned}
\end{equation}
with
\begin{equation}
g_\pm(t)=\frac12\left(e^{-im_{\rm H}t-\frac12\Gamma_{\rm H}t}\pm e^{-im_{\rm L}t-\frac12\Gamma_{\rm L}t}\right).
\end{equation}
A particularly useful quantity that measures the level of CP and baryon number
violation is the quantity,
\begin{equation}
A_{\cal B}=\frac{P_{{\cal B}\to{\cal B}}-P_{{\cal B}\to\bar{\cal B}}+P_{\bar{\cal B}\to{\cal B}}-P_{\bar{\cal B}\to\bar{\cal B}}}{P_{{\cal B}\to{\cal B}}+P_{{\cal B}\to\bar{\cal B}}+P_{\bar{\cal B}\to{\cal B}}+P_{\bar{\cal B}\to\bar{\cal B}}},
\end{equation}
where, e.g., $P_{{\cal B}\to\bar{\cal B}}$ is the time integrated probability
for an intitial $\cal B$ state to oscillate into a $\bar{\cal B}$ and the other
terms are defined analogously. In terms of the elements of $\cal H$, this can
be concisely expressed,
\begin{equation}
A_{\cal B}=\frac{2{\rm Im}\left(M_{12}^\ast\Gamma_{12}\right)}{\Gamma_{\cal B}^2+4\left|M_{12}\right|^2}.
\label{eq:AB}
\end{equation}
This expresses the familiar fact that CP violation requires a phase difference
between the absorptive and dispersive parts of the transition amplitudes.

The dispersive part of the transition amplitude, $M_{12}$, is dominantly given
by off-shell $\chi_i$ exchange in our model, as seen in Fig.~\ref{fig:six
quark_chi_phi}. We have already written down what we need to estimate this in
Eqs.~(\ref{eq:CBB}) and (\ref{eq:delta}), resulting in
\begin{equation}
M_{12}\sim\kappa^2\sum_{i}\frac{{m_{\chi}}_i}{m_{\cal B}^2-{m_{\chi}}_i^2}\left(\frac{g_{ud}^\ast y_{id^\prime}}{m_\phi^2}\right)^2.
\label{eq:M12}
\end{equation}
Here, $u$, $d$, and $d^\prime$ refer to the flavors that comprise $\cal B$ and
we have assumed, if $d\neq d^\prime$, that $g_{ud}^\ast y_{id^\prime}\gg
g_{ud^\prime}^\ast y_{id}$. If we concentrate on the contribution due to a
particular $\chi_i$ and express it in terms of its mass difference from the
baryon, $\Delta m_{{\cal B}i}=m_{\cal B}-m_{\chi_i}$, we have
\begin{equation}
\begin{aligned}
\left|M_{12}\right|_i&\sim\frac{\kappa^2}{2\Delta m_{{\cal B}i}}\left|\frac{g_{ud}^\ast y_{id^\prime}+g_{ud^\prime}^\ast y_{id}}{m_\phi^2}\right|^2
\\
&\simeq 8\times10^{-16}~{\rm GeV}\left(\frac{500~\rm MeV}{\Delta m_{{\cal B}i}}\right)
\\
&\quad\times\left(\frac{600~\rm GeV}{m_\phi/\sqrt{\left|g_{ud}^\ast y_{id^\prime}+g_{ud^\prime}^\ast y_{id}\right|}}\right)^4.
\end{aligned}
\label{eq:M12i}
\end{equation}

The absorptive part of the transition amplitude requires an on-shell state into
which both $\cal B$ and $\bar{\cal B}$ can decay. This requires at least
$\chi_1$ to be light enough for either baryon or antibaryon to decay into it.
CP violation will be largest when the mass splitting between $\chi_1$ and $\cal
B$ is not too large. In this case the most important states for $\Gamma_{12}$
are decays of $\cal B$ to $\chi_1$ plus a meson. The contribution from
$\chi_1\pi^0$, for instance, can be estimated using the effective Lagrangian,
\begin{equation}
{\cal L}_{\rm eff}\supset -y_{i{\cal B}}\pi^0\bar{\cal B}i\gamma^5\chi_i+{\rm h.c.},
\end{equation}
where
\begin{equation}
y_{i{\cal B}}\sim \frac{4\pi\kappa}{m_{\cal B}}\frac{g_{ud}^\ast y_{id^\prime}}{m_\phi^2}.
\end{equation}
The factor of $4\pi$ in this expression accounts for the nonperturbative nature
of the interaction, which is similar to the pion-nucleon vertex. This
interaction gives a contribution to $\Gamma_{12}$ of
\begin{equation}
\Gamma_{12}\sim \sum_i\frac{y_{i{\cal B}}^2m_{\chi_i}}{32\pi}\left(1+r_{\chi_i}-r_{\pi^0}\right)\lambda^{1/2}\left(1,r_{\chi_i},r_{\pi^0}\right),
\end{equation}
where $r_{\chi_i,\pi^0}=m_{\chi_i,\pi^0}^2/m_{\cal B}^2$ and
$\lambda\left(a,b,c\right)=a^2+b^2+c^2-2ab-2ac-2bc$. The magnitude of the
contribution to $\Gamma_{12}$ from a particular $\chi_i$ is roughly
\begin{equation}
\begin{aligned}
\left|\Gamma_{12}\right|_i&\sim \frac{y_{i{\cal B}}^2}{8\pi}\Delta m_{{\cal B}i}\sim \frac{2\pi\kappa^2}{m_{\cal B}^2}\left|\frac{g_{ud}^\ast y_{id^\prime}}{m_\phi^2}\right|^2
\\
&\simeq 1\times10^{-16}~{\rm GeV}\left(\frac{\Delta m_{{\cal B}i}}{500~\rm MeV}\right)
\\
&\quad\times\left(\frac{5~\rm GeV}{m_{\cal B}}\right)^2\left(\frac{600~\rm GeV}{m_\phi/\sqrt{\left|g_{ud}^\ast y_{id^\prime}\right|}}\right)^4.
\end{aligned}
\end{equation}
We now see the reason we need at least two Majorana fermions. If there were
only a single Majorana fermion, $\chi_1$, that contributed to $M_{12}$ and
$\Gamma_{12}$, they would have the same phase and $A_{\cal B}$ in
Eq.~(\ref{eq:AB}) would vanish. Thus, we use contributions from $\chi_1$ and
$\chi_2$ exchange to obtain a physical, CP-violating phase difference between
$M_{12}$ and $\Gamma_{12}$.

The ratio of the single meson contribution to $\Gamma_{12}$ from $\chi_1$ to
its contribution to $M_{12}$ is
\begin{equation}
\begin{aligned}
\left|\frac{\Gamma_{12}}{M_{12}}\right|_1&\sim 4\pi\left(\frac{\Delta m_{{\cal B}1}}{m_{\cal B}}\right)^2
\\
&\simeq0.1\left(\frac{\Delta m_{{\cal B}1}}{500~\rm MeV}\right)^2\left(\frac{5~\rm GeV}{m_{\cal B}}\right)^2.
\end{aligned}
\end{equation}
The CP-violating quantity $A_{\cal B}$ in Eq.~(\ref{eq:AB}) linearly depends on
$\left|\Gamma_{12}\right|$. Without finely tuning the contributions due to
$\chi_1$ and $\chi_2$ against each other, this value of the ratio due to
$\chi_1$ alone is roughly as large as the total ratio
$\left|\Gamma_{12}/M_{12}\right|$ can get.

\section{Collider Constraints}
\label{sec:colliders}
To obtain a large amount of CP violation in heavy baryon oscillation, it will 
be clear that the lightest two Majorana fermions must have masses on the order 
of a few GeV along with couplings to quarks that are not too small. In this
discussion, we consider the two lightest Majorana fermions. The third, $\chi_3$
must be weakly coupled in the minimal version of the model, due to cosmological
considerations as we will see in Sec.~\ref{sec:cosmo}. 

The constraints that we will discuss in this section require that $\phi$ have a
mass of at least a few hundred $\rm GeV$. In this case, $\chi_i$ decays can be
analyzed by integrating out the scalar in Eq.~(\ref{eq:L_int}) resulting in
four-fermion interactions,
\begin{equation}
-\frac{g_{ud}y_{id^\prime}}{m_\phi^2}\bar \chi_i \bar{u}_R d_R^c d^{\prime c}_R+{\rm h.c.}
\label{eq:4fermi_chi}
\end{equation}
(For $d\neq d^\prime$, we have again assumed that $g_{ud}y_{id^\prime}\gg
g_{ud^\prime}y_{id}$.) Both the interactions responsible for the decay of the
Majorana fermions and those that source CP-violating baryon oscillations are of
the same form. The quarks involved in the decay operator must be lighter than
those responsible for baryon oscillations and the couplings responsible for
decay must be relatively smaller, to avoid stronger dinucleon decay limits from
$\Delta B=2$ quarks involving light quarks.

The interaction in Eq.~(\ref{eq:4fermi_chi}) allows for the decay $\chi_i\to
udd^\prime$, where $u$, $d$, and $d^\prime$ are up- and down-type quarks light
enough for this to be kinematically allowed. It is reasonable to assume that
one mode dominates their allowed branchings and in this case, their lifetimes
are
\begin{equation}
\begin{aligned}
\tau_{\chi_{i}}&\sim\frac{2\left(8\pi\right)^3}{m_{\chi_{1,2}}^5}\left|\frac{m_\phi^2}{g_{ud}\,y_{{i}d^\prime}}\right|^2
\\
&\simeq 10^{-6}~{\rm s}\left(\frac{5~\rm GeV}{m_{\chi_{i}}}\right)^5\left(\frac{m_\phi/\sqrt{g_{ud}\,y_{{i}d^\prime}}}{20~\rm TeV}\right)^4.
\end{aligned}
\label{eq:chi_lifetime}
\end{equation}
For $m_{\chi_i}=5~\rm GeV$, with couplings $g_{ud}\,y_{{i}d^\prime}\lesssim
(m_\phi/20~\rm TeV)^2$ the $\Delta B=2$ transition amplitude in the $u d
d^\prime$ system is less than $10^{-22}~\rm GeV$, avoiding conflict with
constraints from dinucleon decay (see Table~\ref{tab:operators}). Furthermore,
if $m_{\chi_i}=5~\rm GeV$, as long as $g_{ud}\,y_{{i}d^\prime}\gtrsim
(m_\phi/350~\rm TeV)^2$, $\tau_{\chi_{i}}\lesssim 0.1~\rm s$ which is a short
enough lifetime to avoid spoiling successful BBN (see,
e.g.,~\cite{Jedamzik:2006xz}).

We might ask whether instead of ensuring that $\chi_{1,2}$ decay fast enough to
avoid spoiling BBN, the lightest fermion $\chi_1$ could instead be long enough
lived to serve as dark matter. First we note that the range allowed for
kinematic stability of both proton and $\chi_{1}$ is extremely fine tuned, with
the $\chi_1$ mass between $m_p-m_e$ and $m_p+m_e$. If we assume all the
$\chi$'s participate in a viable heavy flavor baryogenesis mechanism, we will
see that we must require the $\chi_{1,2}$ masses to be around 3--5 GeV, and we
also need sufficiently large four-fermion interactions involving $\chi_{1,2}$
and heavy flavor quarks, suppressed by a scale $\Lambda_{\rm heavy}\equiv
m_\phi/\sqrt{g y}\sim 600~\rm GeV$. Here $g$ and $y$ here label couplings with
the relevant flavor structure. Based on our discussion in
Sec.~\ref{sec:Heavy_Flavor_Baryon_Osc}, at the one loop level we must generate
four-fermion operators involving light quarks (into which $\chi_{1,2}$ can
decay) are generated with a scale $\Lambda_{\rm light}\gtrsim 10^4 \Lambda_{\rm
heavy}\sim 10^6~\rm GeV$. This provides a lower bound on the strength of the
light quark four-fermion operator which gives an upper bound on the
$\chi_{1,2}$ lifetime of $\tau_{\chi{1,2}}\lesssim 1000~\rm s$ in the absence
of fine-tuning against some other source of this operator. There is also an
unavoidable decay channel that comes from the mixing of $\chi_{1,2}$ with the
heavy flavor baryon, $\cal B$, whose oscillations are responsible for the BAU,
with a mixing angle $\theta\sim\left|M_{12}\right|/\Delta m$ where $\Delta m$
is the mass splitting between the Majorana fermions and $\cal B$. This mixing
leads to the decay of $\chi_{1,2}$ into $\cal B$'s decay channels with a
partial width proportional to $\theta^2\Gamma_{\cal B}$, which is much shorter
than the lifetime of the universe. 

We see that the Majorana fermions are generically unstable but long-lived on
the scale of collider experiments and appear as missing energy. Decay lengths
on the order of $10^2$ to $10^7~\rm m$ are expected, potentially relevant for
the recently proposed MATHUSLA detector~\cite{Chou:2016lxi} which is optimized
to search for long-lived particles. In what follows, to analyze collider
constraints on the new scalar $\phi$ we will assume that any $\chi_i$ produced
at a collider is invisible and defer discussion of the displaced decay
signatures at, e.g., MATHUSLA.

Now that we know that the $\chi_i$'s are invisible at colliders, we can
understand how the scalars appear when produced in hadron collisions. Because
$\phi$ is a color fundamental, if it is kinematically accessible,
$\phi\phi^\ast$ pairs are easily produced in proton-(anti)proton collisions,
and the signatures are essentially those of squarks in RPV SUSY. In addition to
QCD production, (single) scalars can be resonantly produced in the presence of
some nonzero $g_{ud}$. Once produced, the scalar decays through one of the
interactions in Eq.~(\ref{eq:L_int}), either to quark pairs with a rate
\begin{equation}
\Gamma_{\phi\to \bar u\bar d}\simeq\sum_{i,j}\frac{\left|g_{u_id_j}\right|^2}{16\pi}m_\phi,
\label{eq:phi_to_jets_rate}
\end{equation}
or to $\chi_i$ plus a quark,
\begin{equation}
\Gamma_{\phi\to\chi d}\simeq\sum_{i,j}\frac{\left|y_{id_j}\right|^2}{16\pi}m_\phi,
\label{eq:phi_to_chi_rate}
\end{equation}
where we have assumed that $m_\phi$ is much larger than the mass of any decay
product. Therefore, these scalars can appear in searches for dijet resonances
(either singly or pair produced) and (mono)jets and missing energy. Which
search is most sensitive depends on $m_\phi$ and the branching fractions for
$\phi\to \bar u \bar d$ and $\phi\to \chi d$.

\begin{figure}
\includegraphics[width=\linewidth]{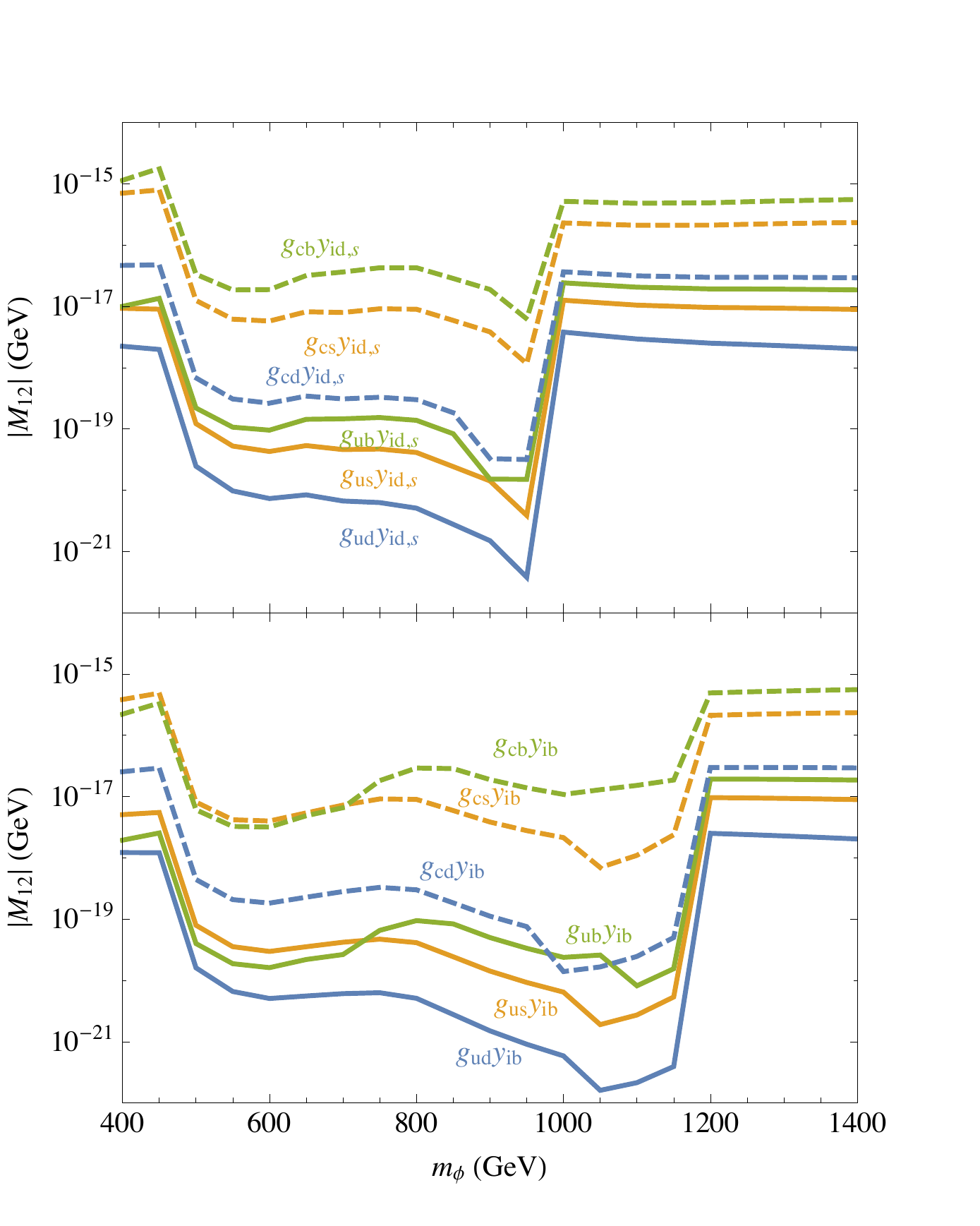}
\caption{Upper limits on $M_{12}$ as functions of $m_\phi$ that result from
collider searches for diject resonances and jets plus missing energy, assuming
the dominance of the product of couplings $g_{ud}y_{i d^\prime}$ indicated,
where $u$ and $d^{(\prime)}$ label generic up- and down-type quarks,
respectively. We have taken $\Delta m_{\cal B}=m_{\cal B}-m_\chi=200\rm \ MeV$.
Top: The limits when $y_{id}$ or $y_{is}$ are dominant. Bottom: The limits when
$y_{ib}$ is dominant. Solid curves show the limits in the case where the charge
$2/3$ quark involved is $u$ while dashed lines show the limit in the case of
the $c$ quark.}
\label{fig:M12coll}
\end{figure}

Taken together, LHC searches for pair produced dijet resonances, both
with~\cite{Khachatryan:2014lpa} and without~\cite{ATLAS:2016sfd} heavy flavor
in the final states, as well as standard SUSY searches for
($b$-tagged~\cite{Aaboud:2016nwl,CMS:2016mwj} or
not~\cite{ATLAS:2016kts,CMS:2016mwj}) jets plus missing energy rule out $\phi$
masses below about $400~\rm GeV$. Above this mass, limits from pair produced
dijet resonances are no longer constraining while resonant production of a
single $\phi$ with a rate proportional to $\left|g_{ud}\right|^2$ for some $u$,
$d$ is important~\cite{Monteux:2016gag}. We use the limits from resonant dijet
production from~\cite{Monteux:2016gag} and recast searches for jets and missing
energy~\cite{Aaboud:2016nwl,CMS:2016mwj,ATLAS:2016kts} as well as
monojets~\cite{Khachatryan:2014rra} to find limits on the couplings $g_{ud}$
and $y_{id^\prime}$ as functions of $m_\phi$. We find this limit for every
flavor $u$, $d$, and $d^\prime$, assuming that only $g_{ud}$ and
$y_{id^\prime}$ are relevant. Given these limits, the maximum value of the
product of couplings $g_{ud}y_{id^\prime}$ at each $m_\phi$ can be found, and
taking a value of the mass splitting between $\chi_i$ and the $udd^\prime$
baryon, which can be turned into an upper limit on the transition amplitude
$\delta_{udd^\prime}=M_{12}$ in the $udd^\prime$ system.\footnote{Note that we
perform this scan for a single Majorana fermion. Including a second, as we must
to obtain CP violation, does not change the allowed values by more than an
${\cal O}(1)$ factor which, given our level of precision, is unimportant. The
third, $\chi_3$ must be more weakly coupled than $\chi_{1,2}$ and can be even
more safely neglected here.} We show the upper limit on $M_{12}$ as a function
of $m_\phi$ for each pattern of flavors $u$, $d$, and $d^\prime$, assuming the
dominance of one particular pair of couplings $g_{ud}$, $y_{id^\prime}$ and a
mass splitting between $m_{\chi_i}$ and the $udd^\prime$ baryon of $200~\rm
MeV$ in Fig.~\ref{fig:M12coll}. We also show the largest value of $M_{12}$
allowed from collider searches in each neutral baryon system in
Table~\ref{tab:operators}.

\section{Hadron phenomenology}
\label{sec:hadrons}
\subsection{Hadron decays}
\label{sec:hadrondecays}
After integrating out the heavy colored scalars, four-fermion interactions
between the Majorana fermions and quarks are generated as in
Eq.~(\ref{eq:4fermi_chi}). These can lead to new decays of hadrons to final
states that differ in baryon number by one unit along with any kinematically
accessible $\chi_i$, e.g.,
\begin{equation}
\begin{aligned}
\rm meson&\to{\rm baryon}+\chi_i\,[+\,{\rm meson(s)}],\\
\rm baryon&\to{\rm meson(s)}+\chi_i.
\end{aligned}
\end{equation}
As we showed in Sec.~\ref{sec:colliders}, on the scale of particle physics
experiments, $\chi_i$ appear as missing energy. 

For definiteness, let us focus now on four-fermion interactions that involve
the $b$ quark and the lightest Majorana fermion. This is potentially relevant
to the case where baryons containing $b$ quarks undergo CP-violating
oscillations in the early Universe, producing the BAU; operators involving
heavy quarks are less constrained by dinucleon decay and are therefore more
promising candidates, cf. Table~\ref{tab:operators}. Similar considerations
apply for operators involving lighter quarks. 

Consider, as a definite example, $b$ decays through the operator
\begin{equation}
-\frac{g_{ub}y_{1d}}{m_\phi^2}\chi_1 u_R d_R b_R,
\end{equation}
where $u$ and $d$ here are the actual up and down quarks. (We have omitted the
contribution to this operator from $g_{ud}y_{1b}$ which is more constrained by
collider searches.) The rate for the $b$ quark to decay through such an
interaction is
\begin{equation}
\begin{aligned}
&\Gamma_{b\to\chi_1\bar u\bar d}\sim\frac{m_b\Delta m^4}{60\left(2\pi\right)^3}\left(\frac{g_{ub}y_{1d}}{m_\phi^2}\right)^2+{\cal O}\left(\frac{\Delta m^5}{m_b^5}\right)
\\
&~\simeq2\times10^{-15}~{\rm GeV}\left(\frac{\Delta m}{2~\rm GeV}\right)^4\left(\frac{1.2~\rm TeV}{m_\phi/\sqrt{g_{ub}y_{1d}}}\right)^4.
\end{aligned}
\label{eq:bdecayrate}
\end{equation}
In this expression $\Delta m$ is the mass splitting between $\chi_1$ and the
bottom quark (we have ignored masses in the final state besides $m_{\chi_1}$).
We have chosen to normalize this expression on values of the mass splitting and
$m_\phi/\sqrt{g_{ub}y_{1d}}$ that result in a transition amplitude of
$\left|M_{12}\right|\sim 10^{-17}~\rm GeV$ in the $\Lambda_b^0=(udb)$ baryon
system, which is the rough collider limit. Given this mass splitting, this can
lead to decays of $B^+$ mesons to a nucleon to $\chi_1$ with a branching ratio
of
\begin{equation}
\begin{aligned}
{\rm Br}_{B^\pm\to N\chi_1+X}&\sim6\times10^{-3}\left(\frac{\Delta m}{2~\rm GeV}\right)^4
\\
&\quad\times\left(\frac{1.2~\rm TeV}{m_\phi/\sqrt{g_{ub}y_{1d}}}\right)^4,
\end{aligned}
\label{eq:bbranching}
\end{equation}
where $X$ represents possible additional pions. This is not a small branching
fraction, although final states of this form have not yet been searched for in
$B$ meson decays. However, the requirement that the final state hadrons carry
baryon number means that this decay is kinematically forbidden if
$m_{\chi_1}>m_{B^\pm}-m_p=4.34~\rm GeV$. Decays of bottom baryons would be
allowed to proceed for splittings down to $m_\pi$, and one could expect
branching ratios on the order of $10^{-3}$ for the parameters in
Eq.~(\ref{eq:bbranching}).\footnote{The calculation of the baryon decay rate to
$\chi_1$ and a single meson is essentially the same as that of $\Gamma_{12}$ in
Sec.~\ref{heavyBosc}, modulo a factor of $m_b/m_{\chi_1}\sim{\cal
O}\left(1\right)$.} 

We also expect ``wrong sign'' decays of heavy baryons in this model, following
a ${\cal B}\to\bar{\cal B}$ oscillation, with a branching fraction that is
roughly
\begin{equation}
\frac12\frac{\left|M_{12}\right|^2}{\Gamma_{\cal B}^2}.
\end{equation}
Consider, e.g., the $\Omega_c^0$. Given the constraints that appear in
Table~\ref{tab:operators}, this branching could potentially be as large as
$10^{-7}$. The Belle II experiment hopes to collect $\sim 50~\rm ab^{-1}$ of
$e^+e^-$ data at $\sqrt s=10.56~\rm GeV$ collecting about $50\times 10^9$ $B$
meson pairs. If $\Omega_c^0$ baryons are produced in $2\%$ of $B$ meson decays
(comparable to the measured production of $\Lambda_c$ baryons), then there
would be a sample of about $10^9$ $\Omega_c^0$'s and $\bar\Omega_c^0$'s. Thus,
there could be a few hundred ``wrong sign'' decays in the data sample. While
this would be a challenging measurement, it is interesting that it is in
principle observable at the next generation $B$-factory given current
experimental limits.

We mention here that baryon-number--violating decays of baryons along these
lines have been searched for by the CLAS
Collaboration~\cite{McCracken:2015coa}. The branching fraction for $\Lambda\to
K^0_S+{\rm inv.}$ is limited to less than $2\times 10^{-5}$ while that for
$\Lambda\to \bar p \pi^+$ must be less than $9\times 10^{-7}$ which are
sensitive to the operator $(uds)^2$. While interesting, these limit
$\delta_{(uds)^2}=\left|M_{12}\right|$ to less than about $10^{-18}~\rm GeV$,
which is less strong than the limit on this operator from null searches for
dinucleon decay. In light of the less stringent limits from dinucleon decay on
operators involving heavy flavor, it would be highly desirable for searches for
$\Delta B=2$ decays of baryons with heavy quarks to be performed.

\subsection{Meson oscillations}
\label{sec:mesonosc}
In addition to the decays described above, the new interactions could lead to
flavor-changing oscillations of neutral mesons. The limits from these processes
on this model were considered in Ref.~\cite{Ghalsasi:2015mxa}; we refer the
reader to~\cite{Ghalsasi:2015mxa} and references therein for further details.

Avoiding these constraints requires a suppression of particular combinations of
flavor-violating couplings. For example, considering Kaon oscillations, given
$m_\phi\gtrsim 400~\rm GeV$, $y_{s1}$ and $y_{s2}$ could be ${\cal O}(1)$
provided $y_{d1},y_{d2}\lesssim 10^{-2}$. Similar considerations apply for
flavor-violating combinations of the couplings $g_{us}$ and $g_{ud}$. The
constraints on charm and bottom couplings from $D$ and $B$ oscillations are
less severe.

From the model building point of view, flavor-changing meson oscillations can
be naturally avoided, e.g. charging the scalar under a symmetry so that
$F_1-F_2$ is conserved, where $F_{1,2}$ label flavor quantum numbers.

\section{Cosmological Production of the Baryon Asymmetry}
\label{sec:cosmo}
We now answer in detail the question of how the baryon asymmetry of the
Universe is produced in this model. In addition to the CP and baryon number
violation described above, a nonzero asymmetry requires a departure from
thermal equilibrium. The simplest possibility for this is to assume that
$\chi_3$ is very weakly coupled. It is therefore long-lived and decays out of
equilibrium, producing the baryons that undergo CP- and $B$-violating
oscillations.

At temperatures below $m_{\chi_3}$, the equations that determine the radiation
and $\chi_3$ energy densities are
\begin{align}
&\frac{d\rho_{\rm rad}}{dt}+4H\rho_{\rm rad}=\Gamma_{\chi_3}\rho_{\chi_3},
\label{eq:rhorad}
\\
&\frac{d\rho_{\chi_3}}{dt}+3H\rho_{\chi_3}=-\Gamma_{\chi_3}\rho_{\chi_3}.
\label{eq:rhomat}
\end{align}
$H$ is the Hubble parameter which is related to the total energy density,
\begin{equation}
H=\sqrt{\frac{8\pi}{3}\frac{\rho}{M_{\rm Pl}^2}}\simeq\sqrt{\frac{8\pi}{3}\frac{\rho_{\rm rad}+\rho_{\chi_3}}{M_{\rm Pl}^2}},
\end{equation}
where $M_{\rm Pl}=1.22\times10^{19}~\rm GeV$ is the Planck mass. In the absence
of $\chi_3$ decays, $\rho_{\rm rad}$ and $\rho_{\chi_3}$ simply redshift like
radiation and matter energy densities, respectively. The right-hand sides of
these equations describe how $\chi_3$ decays cause the energy density in matter
to decrease while dumping energy into the plasma.

In addition to depositing energy in the plasma, some of the $\chi_3$ decays
produce baryons and antibaryons, $\cal B$ and $\bar{\cal B}$, that can
oscillate and decay, violating CP and $B$. For this to occur, the temperature
of the Universe needs to be below the QCD confinement temperature, $T_{\rm
QCD}\simeq 200~\rm MeV$. On the timescale of the expansion of the Universe,
$H^{-1}$, the \mbox{(anti-)baryons} produced this way rapidly oscillate and
decay, producing a net $B$ asymmetry. However, because of the presence of the
plasma, with which they can interact, as well as their large annihilation cross
section, ${\cal B}\leftrightarrow\bar{\cal B}$ can decohere in this
environment, suppressing the asymmetry that is generated. Properly accounting
for this requires a density matrix treatment, which has been used in a
cosmological context for neutrino oscillations and oscillating asymmetric dark
matter~\cite{Cirelli:2011ac,Tulin:2012re}. Following Ref.~\cite{Tulin:2012re}
(see~\cite{Ipek:2016bpf} for a similar analysis in the context of
baryogenesis-related oscillations), we can write the Boltzmann equations that
govern the evolution of the number density of the $\cal B$-$\bar{\cal B}$
system,
\begin{equation}
\begin{aligned}
\frac{dn}{dt}+3Hn&=-i\left({\cal H}n-n{\cal H}^\dagger\right)-\frac{\Gamma_\pm}{2}\left[O_\pm,\left[O_\pm,n\right]\right]
\\
&\quad-\langle\sigma v\rangle_\pm\left(\frac12\left\{n,O_\pm\bar n O_\pm\right\}-n_{\rm eq}^2\right)
\\
&\quad+\frac12 \frac{\Gamma_{\chi_3}\rho_{\chi_3}}{m_{\chi_3}}{\rm Br}_{\chi_3\to{\cal B}}O_+,
\end{aligned}
\label{eq:dmevo}
\end{equation}
where the last term describes $\cal B$ and $\bar{\cal B}$ production through
$\chi_3$ decay. ${\rm Br}_{\chi_3\to{\cal B}}$ is the branching ratio for
$\chi_3$ to decay to $\cal B$ or $\bar{\cal B}$. In this equation $n$ and $\bar
n$ are density matrices,
\begin{equation}
n=\left(\begin{array}{cc}
n_{{\cal B}{\cal B}} & n_{{\cal B}\bar{\cal B}} \\
n_{\bar{\cal B}{\cal B}} & n_{\bar{\cal B}\bar{\cal B}}
\end{array}\right),~\bar n=\left(\begin{array}{cc}
n_{\bar{\cal B}\bar{\cal B}} & n_{{\cal B}\bar{\cal B}} \\
n_{\bar{\cal B}{\cal B}} & n_{{\cal B}{\cal B}}
\end{array}\right),
\end{equation}
and $n_{\rm eq}$ is the equilibrium density of baryons plus antibaryons. $\cal
H$ is the Hamiltonian seen in Eq.~(\ref{eq:ham}). $\langle\sigma v\rangle_\pm$
and $\Gamma_\pm$ are thermally-averaged annihilation cross sections and
scattering rates on the plasma, respectively. $O_\pm$ is a matrix
\begin{equation}
O_\pm=\left(\begin{array}{cc}
1 & 0 \\
0 & \pm 1
\end{array}\right).
\end{equation}
The subscript of $\langle\sigma v\rangle_\pm$ and $\Gamma_\pm$, i.e. whether
they appear with $O_+$ or $O_-$ in Eq.~(\ref{eq:dmevo}), is determined by the
behavior of the effective Lagrangian that gives rise to these interactions
under charge conjugation of {\it only} the heavy baryons, ${\cal
B}\leftrightarrow\bar{\cal B}$, ${\cal L}_{\rm eff}\leftrightarrow\pm{\cal
L}_{\rm eff}$. Interactions that do not change sign are said to be flavor-blind
while those that do are flavor-sensitive. For example, $\cal B$ and $\bar{\cal
B}$ can scatter on light charged particles in the plasma through their magnetic
moment, $\mu$, which corresponds to a term in the effective Lagrangian of
\begin{equation}
\frac{i\mu}{4}\bar{\cal B}\left[\gamma^\nu,\gamma^\rho\right]{\cal B}F_{\nu\rho}.
\label{eq:magmom}
\end{equation}
Under ${\cal B}\leftrightarrow\bar{\cal B}$ this term changes sign, so the rate
for scattering via the magnetic moment appears with $O_-$ in the Botzmann
equation. 

It is useful to work in terms of the quantities
\begin{equation}
\begin{aligned}
\Sigma&\equiv n_{{\cal B}{\cal B}}+n_{\bar{\cal B}\bar{\cal B}},~\Delta\equiv n_{{\cal B}{\cal B}}-n_{\bar{\cal B}\bar{\cal B}},
\\
\Xi&\equiv n_{{\cal B}\bar{\cal B}}-n_{\bar{\cal B}{\cal B}},~\Pi\equiv n_{{\cal B}\bar{\cal B}}+n_{\bar{\cal B}{\cal B}}.
\end{aligned}
\end{equation}
In this basis the Boltzmann equations are
\begin{equation}
\begin{aligned}
\left(\frac{d}{dt}+3H\right)\Sigma&=\frac{\Gamma_{\chi_3}\rho_{\chi_3}}{m_{\chi_3}}{\rm Br}_{\chi_3\to{\cal B}}-\Gamma_{\cal B}\Sigma
\\
&\quad-\left({\rm Re}\,\Gamma_{12}\right)\Pi+i\left({\rm Im}\,\Gamma_{12}\right)\Xi
\\
&\quad-\frac12\Big[\big(\langle\sigma v\rangle_++\langle\sigma v\rangle_-\big)\left(\Sigma^2-\Delta^2-4n_{\rm eq}^2\right)
\\
&\quad\quad\quad+\big(\langle\sigma v\rangle_+-\langle\sigma v\rangle_-\big)\left(\Pi^2-\Xi^2\right)\Big],
\\
\left(\frac{d}{dt}+3H\right)\Delta&=-\Gamma_{\cal B}\Delta+2i\left({\rm Re}\,M_{12}\right)\Xi+2\left({\rm Im}\,M_{12}\right)\Pi,
\\
\left(\frac{d}{dt}+3H\right)\Xi&=-\big(\Gamma_{\cal B}+2\Gamma_-+\langle\sigma v\rangle_+\Sigma\big)\Xi
\\
&\quad\quad+2i\left({\rm Re}\,M_{12}\right)\Delta-i\left({\rm Im}\,\Gamma_{12}\right)\Sigma,
\\
\left(\frac{d}{dt}+3H\right)\Pi&=-\big(\Gamma_{\cal B}+2\Gamma_-+\langle\sigma v\rangle_+\Sigma\big)\Pi
\\
&\quad\quad-2\left({\rm Im}\,M_{12}\right)\Delta-\left({\rm Re}\,\Gamma_{12}\right)\Sigma.
\end{aligned}
\label{eq:Bsys}
\end{equation}
Coherent oscillations from a flavor-symmetric state to an asymmetric state
proceed through $\Sigma\rightarrow\Xi,\Pi\rightarrow\Delta$. Flavor-sensitive
scattering and flavor-blind annihilation suppress $\Xi$ and $\Pi$ and therefore
lead to decoherence.

When they decay, $\cal B$ and $\bar{\cal B}$ create states that carry baryon
number. The flavor-asymmetric configuration contributes to the difference
between the baryon and antibaryon number densities,
\begin{equation}
\left(\frac{d}{dt}+3H\right)\left(n_B-n_{\bar B}\right)=\Gamma_{\cal B}\Delta.
\label{eq:aymm_evolution}
\end{equation}

The dominant interaction of $\cal B$ and $\bar{\cal B}$ with the plasma is
scattering on charged particles (mostly electrons at $T\lesssim100~\rm MeV$)
via the magnetic moment term in Eq.~(\ref{eq:magmom}). The cross section for
this at temperatures well below $m_{\cal B}$ is
\begin{equation}
\frac{d\sigma_{\rm sc}}{d\Omega}=\alpha^2\mu^2\left(\frac{1+\sin^2\theta/2}{\sin^2\theta/2}\right)
\end{equation}
which diverges at small scattering angle, $\theta\to 0$. This divergence is cut
off at finite temperature by the inverse photon screening length, $m_\gamma$.
Using this, the total cross section can be estimated as
\begin{equation}
\sigma_{\rm sc}\sim4\pi\alpha^2\mu^2\log\left(\frac{4E^2}{m_\gamma^2}\right)
\end{equation}
where $E$ is the electron energy. Taking $E\sim T$, $m_\gamma\sim e
T/3$~\cite{Thoma:1995ju} and $\mu\sim 1/(2m_{\cal B})$, this gives
\begin{equation}
\sigma_{\rm sc}\sim\frac{\pi\alpha^2}{m_{\cal B}^2}\log\left(\frac{9}{\pi\alpha}\right).
\end{equation}
The (flavor-sensitive) scattering rate is therefore
\begin{equation}
\begin{aligned}
\Gamma_-&=\Gamma_{\rm sc}\sim\sigma_{\rm sc}\left(n_{e^-}+n_{e^+}\right)
\\
&\sim\frac{\pi\alpha^2}{m_{\cal B}^2}\log\left(\frac{9}{\pi\alpha}\right)\times\frac{3\zeta(3)}{\pi^2}T^3
\\
&\sim 10^{-11}~{\rm GeV}\left(\frac{5~\rm GeV}{m_{\cal B}}\right)^2\left(\frac{T}{10~\rm MeV}\right)^3.
\end{aligned} 
\label{eq:gammam}
\end{equation}
At temperatures above a few MeV, as is needed for BBN, this rate is larger than
a typical heavy baryon width and therefore strongly affects the $\cal
B$-$\bar{\cal B}$ oscillations.

When solving the Boltzmann equations, we take an annihilation cross section
that is similar to that for $p\bar p$ annihilation at low energies,
\begin{equation}
\langle\sigma v\rangle_++\langle\sigma v\rangle_-=400~{\rm mb}.
\end{equation}
We will find that only the total annihilation cross section and not whether it
is flavor-blind or -sensitive is important, since $\Sigma\gg\Delta,\Xi,\Pi$.
Furthermore the annihilation rate is always much smaller than the scattering
rate at temperatures we are interested in, so its effect on the final asymmetry
is subdominant and can generally be ignored.

\subsection{Sudden Decay Approximation}
\label{sec:suddendecay}
Having removed the heavy baryons from the problem due to the short timescales
in their system, the evolution equations are Eqs.~(\ref{eq:rhorad}),
(\ref{eq:rhomat}), and (\ref{eq:nBsimple}). These involve only the radiation
energy density, $\chi_3$ density, and the baryon asymmetry. They can be simply
studied using a sudden decay approximation to gain a rough estimate of the
baryon asymmetry. We outline this estimate below.

At some high temperature above $m_{\chi_3}$, we assume that $\chi_3$ was in
thermal equilibrium with the plasma, fixing its number density for $T\lesssim
m_{\chi_3}$ to roughly
\begin{equation}
n_{\chi_3}\simeq\frac34\frac{\zeta(3)}{\pi^2}T^3.
\end{equation}
As the Universe cools the energy density in $\chi_3$ and radiation are equal.
This occurs at the temperature
\begin{equation}
T_{\rm eq}=\frac{45\zeta(3)}{2\pi^4g_\ast(T_0)}m_{\chi_3}.
\end{equation}
$g_\ast$ is the effective number of relativistic degrees of freedom and here it is evaluated at $T_0\gtrsim m_{\chi_3}$. This corresponds to the time
\begin{equation}
\begin{aligned}
t_{\rm eq}&=\sqrt{\frac{45}{16\pi^3g_\ast(T_{\rm eq})}}\frac{M_{\rm Pl}}{T_{\rm eq}^2}
\\
&=\frac{1}{\sqrt{5\pi g_\ast(T_{\rm eq})}}\frac{\pi^7g_\ast(T_0)^2}{135\zeta(3)^2}\frac{M_{\rm Pl}}{m_{\chi_3}^2}.
\end{aligned}
\end{equation}
After this the Universe is matter dominated and the energy density in radiation and $\chi_3$ redshift as
\begin{equation}
\rho_{\rm rad}=\frac12\rho_{\rm eq}\left(\frac{t_{\rm eq}}{t}\right)^{8/3},~\rho_{\chi_3}=\frac12\rho_{\rm eq}\left(\frac{t_{\rm eq}}{t}\right)^2.
\end{equation}
We then assume that all of the $\chi_3$'s decay at the time $t_{\rm
dec}=1/\Gamma_{\chi_3}$. The ratio of the energy densities just before decay is
\begin{equation}
\begin{aligned}
\xi&\equiv\frac{\rho_{\chi_3}(t_{\rm dec}^-)}{\rho_{\rm rad}(t_{\rm dec}^-)}=\left(t_{\rm eq}\Gamma_{\chi_3}\right)^{-2/3}
\\
&=15\left[\frac{g_\ast(T_{\rm eq})}{g_\ast(T_0)}\right]^{1/3}\left[\frac{50}{g_\ast(T_0)}\right]
\\
&\quad\quad\times\left(\frac{m_{\chi_3}}{10~\rm GeV}\right)^{4/3}\left(\frac{10^{-22}~\rm GeV}{\Gamma_{\chi_3}}\right)^{2/3}.
\end{aligned}
\end{equation}
We use $t_{\rm dec}^-$ here to indicate the time infinitesimally before decay. 

The dominance of the scattering rate over other scales in the problem allows us
to make some simplifications of the evolution equations that are useful here.
In this limit we can ignore the Hubble rate as well as annihilation and the
equations governing $\cal B$ and $\bar{\cal B}$ in~(\ref{eq:Bsys}) can be
integrated. This results in the evolution equation for the difference between
baryon and antibaryon densities, Eq.~(\ref{eq:aymm_evolution}), becoming
\begin{align}
\left(\frac{d}{dt}+3H\right)\left(n_B-n_{\bar B}\right)&=\frac{\Gamma_{\chi_3}\rho_{\chi_3}}{m_{\chi_3}} \nonumber
\\
&\times\frac{2{\rm Im}\left(M_{12}^\ast\Gamma_{12}\right){\rm Br}_{\chi_3\to{\cal B}}}{\Gamma_{\cal B}\left(\Gamma_{\cal B}+2\Gamma_-\right)+4\left|M_{12}\right|^2} \nonumber
\\
&\simeq\frac{\Gamma_{\chi_3}\rho_{\chi_3}}{m_{\chi_3}}\frac{\Gamma_{\cal B}}{2\Gamma_-}\epsilon,
\label{eq:nBsimple}
\end{align}
which is valid for the cases we consider with
$\left|M_{12}\right|\ll\Gamma_{\cal B}\ll\Gamma_-$. We have defined
\begin{equation}
\epsilon\equiv\frac{2{\rm Im}\left(M_{12}^\ast \Gamma_{12}\right)}{\Gamma_{\cal B}^2}{\rm Br}_{\chi_3\to{\cal B}}\simeq A_{\cal B}{\rm Br}_{\chi_3\to{\cal B}},
\end{equation}
with $A_{\cal B}$ from Eq.~(\ref{eq:AB}).

Using $\epsilon$, we can then relate the baryon asymmetry to the $\chi_3$
number density at decay,
\begin{equation}
\begin{aligned}
\eta_B&=\frac{n_B-n_{\bar B}}{s(t_{\rm dec}^+)}=\frac{n_{\chi_3}(t_{\rm dec}^-)}{s(t_{\rm dec}^-)}\left[\frac{T(t_{\rm dec}^-)}{T(t_{\rm dec}^+)}\right]^3\frac{\Gamma_{\cal B}}{2\Gamma_-}\epsilon
\\
&=\frac34\frac{T(t_{\rm dec}^-)}{m_{\chi_3}}\xi\left[\frac{T(t_{\rm dec}^-)}{T(t_{\rm dec}^+)}\right]^3\frac{\Gamma_{\cal B}}{2\Gamma_-}\epsilon.
\end{aligned}
\end{equation}
Here, $t_{\rm dec}^+$ is the time just after decay. The ratio of the
temperatures just before and after decay is determined by $\rho_{\chi_3}(t_{\rm
dec}^+)=(1+\xi)\rho_{\chi_3}(t_{\rm dec}^-)$ so that
\begin{equation}
\begin{aligned}
\frac{T(t_{\rm dec}^-)}{T(t_{\rm dec}^+)}=(1+\xi)^{-1/4}\simeq \xi^{-1/4},
\end{aligned}
\end{equation}
and
\begin{equation}
\begin{aligned}
\eta_B&\simeq\frac34\frac{\xi^{1/4}T(t_{\rm dec}^-)}{m_{\chi_3}}\frac{\Gamma_{\cal B}}{2\Gamma_-}\epsilon.
\end{aligned}
\end{equation}
The temperature just before decay can be arrived at by evolving the radiation
energy density, resulting in
\begin{equation}
\begin{aligned}
\eta_B&\simeq\frac38\sqrt{\frac{3}{\pi}}\left[\frac{5}{2\pi g_\ast(T_{\rm dec})}\right]^{1/4}\frac{\sqrt{M_{\rm Pl}\Gamma_{\chi_3}}}{m_{\chi_3}}\frac{\Gamma_{\cal B}}{2\Gamma_-}\epsilon.
\end{aligned}
\end{equation}
Using the expression for the scattering rate in Eq.~(\ref{eq:gammam}) evaluated
at $T(t_{\rm dec}^+)$,
\begin{align}
&\eta_B\simeq\frac{\pi^3}{3\zeta(3)}\sqrt{\frac{\pi g_\ast(T_{\rm dec})}{10}}\frac{\Gamma_{\cal B}\epsilon}{\sigma_{\rm sc} m_{\chi_3}\Gamma_{\chi_3}M_{\rm Pl}} \nonumber
\\
&\approx9\times10^{-11}\left[\frac{g_\ast(T_{\rm dec})}{50}\right]^{1/2}\left(\frac{m_{\cal B}}{5~\rm GeV}\right)^2\left(\frac{\Gamma_{\cal B}}{10^{-13}~\rm GeV}\right) \nonumber
\\
&\quad\quad\times\left(\frac{8~\rm GeV}{m_{\chi_3}}\right)\left(\frac{10^{-22}~\rm GeV}{\Gamma_{\chi_3}}\right)\left(\frac{\epsilon}{10^{-5}}\right).
\label{eq:sda}
\end{align}
Therefore we see that a baryon asymmetry of the required size is possible for a
heavy baryon system with $\epsilon\sim10^{-5}$, which requires
$\left|M_{12}\right|/\Gamma_{\cal B}\sim 10^{-2}$ with
$\left|M_{12}\right|/\left|\Gamma_{12}\right|$ not small.

\subsection{Full Solution of the Boltzmann Equations}
\label{sec:fullsoln}
\begin{figure}
\includegraphics[width=\linewidth]{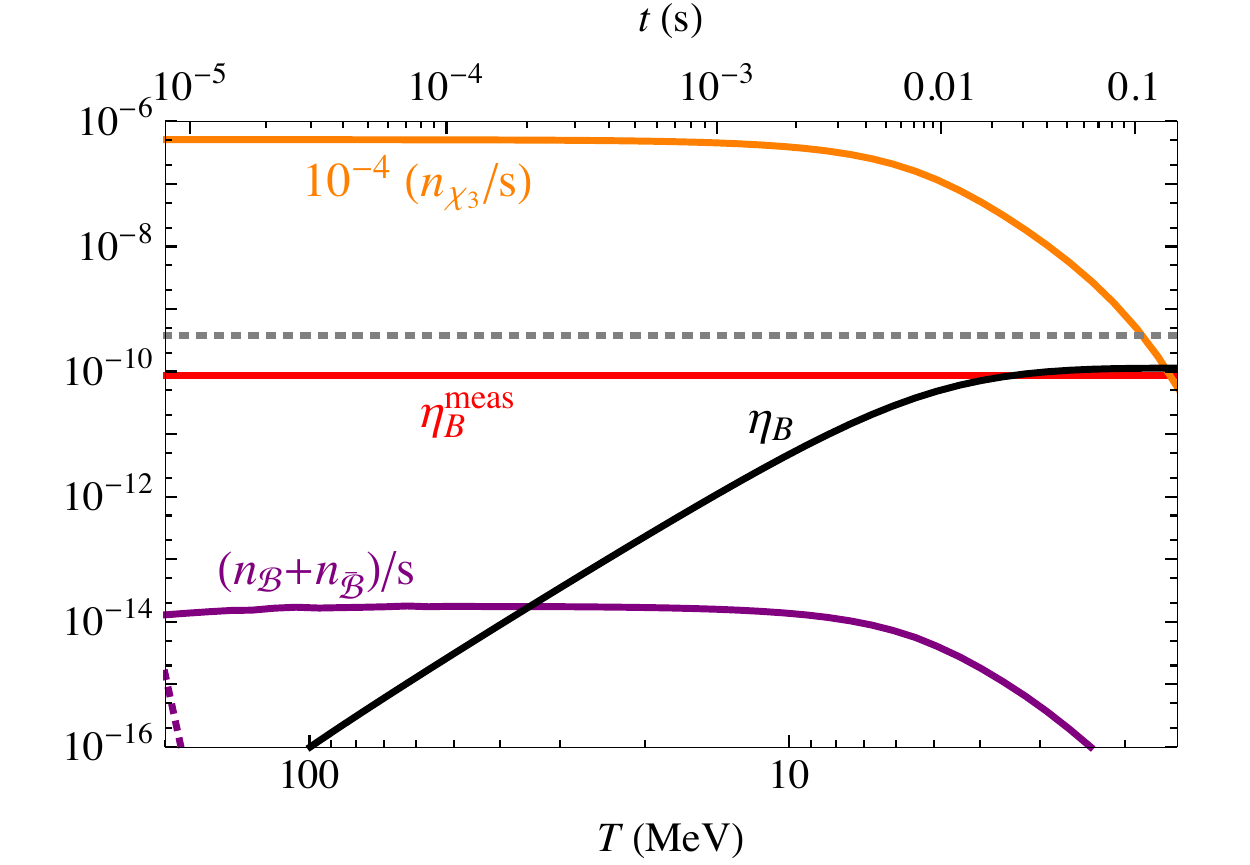}
\caption{$\eta_B=(n_B-n_{\bar B})/s$ (solid, black) as a function of the
temperature or time from a numerical solution of Eqs.~(\ref{eq:rhorad}),
(\ref{eq:rhomat}), and (\ref{eq:Bsys}) for parameters relevant to the
$\Omega_{cb}^0$-$\bar\Omega_{cb}^0$ system: $m_{\cal B}=7~\rm GeV$,
$\Gamma_{\cal B}=3\times 10^{-12}~\rm GeV$,
$\left|M_{12}\right|=3\times10^{-15}~\rm GeV$,
$\left|\Gamma_{12}/M_{12}\right|=0.3$, $\arg(M_{12}^\ast\Gamma_{12})=\pi/2$,
$m_{\chi_3}=7.5~\rm GeV$, $\Gamma_{\chi_3}=3\times 10^{-23}~\rm GeV$, and ${\rm
Br}_{\chi_3\to{\cal B}}=0.35$. We have taken the rate for heavy baryon
scattering on the plasma from Eq.~(\ref{eq:gammam}) and the annihilation cross
section to be $400~\rm mb$. This can be compared against the value of $\eta_B$
(solid, gray) from a solution of Eqs.~(\ref{eq:rhorad}), (\ref{eq:rhomat}), and
(\ref{eq:nBsimple}) as well as using the sudden decay approximation (dashed,
gray) in Eq.~(\ref{eq:sda}). Also shown are the ratio of the number density of
$\chi_3$ to the entropy density (multiplied by $10^{-4}$, solid, orange) and
the ratio of the $\cal B$ plus $\bar{\cal B}$ number densities to the entropy
density (solid, purple). The dashed purple line shows the equilibrium $\cal B$
and $\bar{\cal B}$ density (in units of the entropy density). The measured
value of $\eta_B=8.8\times10^{-11}$ is given by the solid red line.}
\label{fig:eta}
\end{figure}

To get a more precise estimate of the baryon asymmetry, we numerically solve
the system in Eqs.~(\ref{eq:rhorad}), (\ref{eq:rhomat}), and (\ref{eq:Bsys}).
As mentioned above, we need $\left|M_{12}\right|/\Gamma_{\cal B}$ to not be
much smaller than around $10^{-2}$. Looking at Table~\ref{tab:operators}, one
potential candidate is the $\Omega_{cb}^0$ where the dominant coupling involves
the operator $(dcb)^2$. In Fig.~\ref{fig:eta} we show the value of $\eta_B$ as
a function of temperature in the case of the asymmetry being sourced by the
$\Omega_{cb}^0$-$\bar\Omega_{cb}^0$ system, taking $m_{\cal B}=7~\rm GeV$,
$\Gamma_{\cal B}=3\times10^{-12}~\rm
GeV$~\cite{Kiselev:2001fw,*Shah:2016vmd,*Wang:2017mqp},
$\left|M_{12}\right|=3\times 10^{-15}~\rm GeV$,
$\left|\Gamma_{12}/M_{12}\right|=0.3$, $\arg(M_{12}^\ast\Gamma_{12})=\pi/2$,
$m_{\chi_3}=7.5~\rm GeV$, $\Gamma_{\chi_3}=3\times 10^{-23}~\rm GeV$, and ${\rm
Br}_{\chi_3\to\cal B}=0.35$. We have used an annihilation cross section of
$400~\rm mb$ (the results do not depend on whether it is flavor-blind or
-sensitive) and the scattering rate given in Eq.~(\ref{eq:gammam}).

\begin{figure}
\includegraphics[width=\linewidth]{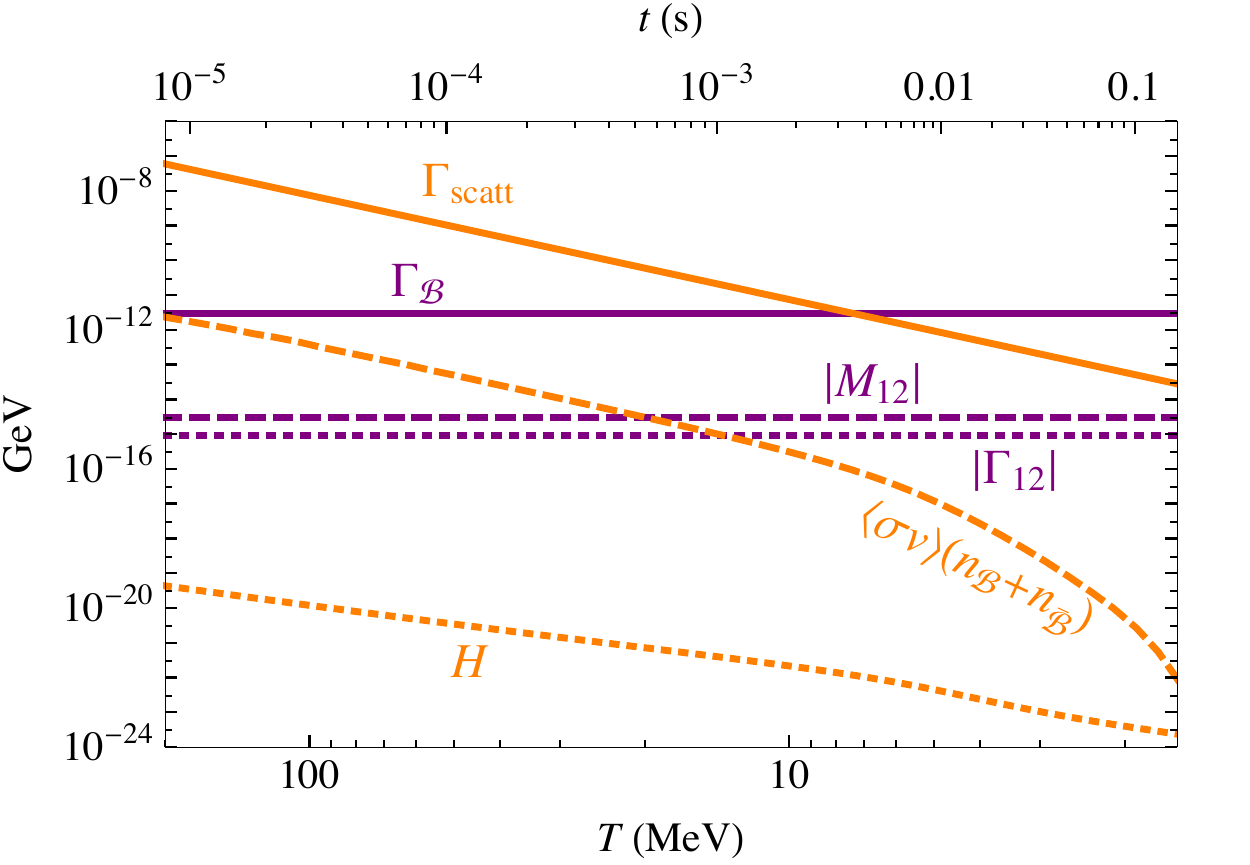}
\caption{The temperature dependence of the rates involved in the numerical
solution of Eqs.~(\ref{eq:rhorad}), (\ref{eq:rhomat}), and (\ref{eq:Bsys}). The
parameters are the same as in Fig.~\ref{fig:eta}. In orange, from top to bottom
are the scattering, annihilation, and Hubble rates. The purple lines indicate
the rates relevant to the $\cal B$-$\bar{\cal B}$ system itself, $\Gamma_{\cal
B}$, $\left|M_{12}\right|$, and $\left|\Gamma_{12}\right|$, from top to bottom,
respectively.}
\label{fig:rates}
\end{figure}

In addition, the temperature dependence of the scattering and annihilation
rates is compared to the expansion rate of the Universe as well as to the rates
governing the baryon-antibaryon system in Fig.~\ref{fig:rates}. As mentioned
before, the (decohering) scattering is the dominant process above temperatures
of about $1~\rm MeV$ and, in particular, is always much larger than the
annihilation rate.

At high temperatures, the heavy baryon density tracks its equilibrium value and
it begins to deviate from its equilibrium value when $\chi_3$'s begin to decay.
Although not directly evident from the plots (except through the change in the
temperature vs. time), the out-of-equilibrium $\chi_3$ particles actually come
to dominate the energy density of the Universe prior to their decay. After the
$\chi_3$ decays, which we assume happens in less than $\sim 0.1 $s, the
Universe undergoes a transition from being matter-dominated to
radiation-dominated, reheating to a temperature above a few MeV.

We have numerically confirmed the rough accuracy of the sudden decay
approximation prediction for $\eta_B$ over much of the parameter space. Maximal
CP violation, and thus more baryon asymmetry per oscillation, occurs for
$\text{arg}\left(M_{12}^{*}\Gamma_{12}\right)=\pi/2$ and larger values of
$\left|M_{12}\right|$ and $\left|\Gamma_{12}\right|$. A larger branching ratio,
$\text{Br}_{\chi_{3}\to\mathcal{B}}$, would produce more oscillating baryons
per Majorana decay. The value of $\eta_B$ that is generated is maximized if
$\chi_3$ decays when the Universe's temperature is about $10~\rm MeV$, i.e.
$\tau_{\chi_3}=1/\Gamma_{\chi_3}\sim 10^{-2}~\rm s$. If it decays earlier than
this, heavy baryon scattering on the plasma leads to decoherence, suppressing
the asymmetry. If if decays later, the Universe does not have a sufficient
baryon asymmetry at the time the neutrinos begin to decouple, when the Universe
is around $3~\rm MeV$.

Given the constraints on the transition amplitudes in
Table~\ref{tab:operators}, the most promising baryon that could allow for a
large enough transition amplitude to source the BAU is the as yet unobserved
$\Omega_{cb}^0$. A relatively large value for $\left|M_{12}\right|$ is needed
in this case, not far from the collider limit, unless ${\rm
Br}_{\chi_3\to\Omega_{cb}^0}$ were rather large. It should be noted that the
collider limits discussed in Sec.~\ref{sec:colliders} which appear in
Table~\ref{tab:operators} depend on the specific model that we considered. It
is conceivable that the model could be extended in a way that makes the
standard collider searches that we considered less constraining. For example,
one could imagine making the $\phi$ decay to a large number of relatively soft
jets by coupling to a heavy vector-like quark and a singlet which decay to a
large number of colored objects. Relaxing these limits could allow for other
heavy flavor baryons to source the BAU, potentially even observed baryons like
the $\Omega_c^0$, $\Lambda_b^0$, and $\Xi_b^0$. On the other hand, since they
involve low-energy effective operators, the dinucleon decay constraints are
less model dependent. Weakening them would require significant tuning of
tree-level operators against those induced by weak interactions.

\section{Summary and Outlook}
\label{sec:summary}
We have presented a model for producing the observed baryon asymmetry of the
Universe which avoids high reheat temperatures. The asymmetry is generated
through CP and $B$-violating oscillations of baryons occurring late in the
hadronization era. Our model minimally introduces three neutral Majorana
fermions and a single colored scalar, and could potentially be embedded into
RPV SUSY. 

The $\Omega_{cb}\sim(scb)$ baryon emerges as our most promising candidate when
constraints due to collider data and dinucleon decay are taken into account.
Note that the constraints from colliders are more model-dependent than those
from the absence of dinucleon decay. Considering only the constraints from
dinucleon decay, additional baryons, e.g., $\Omega_c^0\sim(ssc)$,
$\Lambda_b^0\sim(udb)$, and $\Xi_b^0\sim(usb)$, become viable candidates for
baryogenesis via their oscillation. An interesting avenue for future work would
be constructing models that are less constrained by collider experiments while
preserving a large baryon oscillation rate.

Interesting signatures of this scenario could be present in the large dataset
of the upcoming Belle II experiment. If the lightest Majorana fermion is
sufficiently light, one possible signature would be decays of heavy flavor
hadrons that violate baryon number and involve missing energy. Additionally
there could be heavy flavor baryons that oscillate into their antiparticles at
potentially measurable rates. Exploring the experimental prospects of this
model at high luminosity, lower energy colliders in more detail will be left
for future work.

Constraints from the LHC and the lack of dinucleon decay observation are quite
important, suggesting the possibility of the detection of a signal in one or
both areas. Dinucleon decays are a more model-independent consequence of this
scenario, and because of the requirement of baryon number violation involving
heavy flavors, it is likely to assume that dinucleon decay to kaons would be
dominant. In the case of the LHC, a particular combination of signals in dijet
resonances (singly and pair produced) along with an excess in jets plus missing
energy should be expected. We should mention in this case that a long-lived
neutral particle, $\chi_1$, that decays hadronically is a generic prediction of
this model. The typical $\chi_1$ decay length is in the range of $10^{2-7}~\rm
m$, which could be well probed by the MATHUSLA detector that was recently
proposed. The signal of a long-lived but unstable particle at this experiment
could help disentangle this scenario from others that lead to excesses in jets
plus missing energy.

\begin{acknowledgments}
We would like to thank Brian Batell, Kristian Hahn, and Ahmed Ismail for useful
conversations. The work of DM is supported by PITT PACC through the Samuel P.
Langley Fellowship. The work of TN is supported by Spanish grants
FPA2014-58183-P and SEV-2014-0398 (MINECO), and PROMETEOII/2014/084
(Generalitat Valenciana). The work of AN and KA was supported in part by the
Department of Energy under grant number DE-SC0011637.
\end{acknowledgments}

\bibliography{HBB}

\end{document}